%% file: main.tex
\documentclass[a4paper,12pt,britishm,texlive=2016]{article}

\RequirePackage{graphicx}
\RequirePackage{subfig}
\RequirePackage{mhchem}
\RequirePackage{url}
\RequirePackage{xcolor}

\newcommand{\TeV}{\ensuremath{\text{Te\kern -0.1em V}}}
\newcommand{\GeV}{\ensuremath{\text{Ge\kern -0.1em V}}}
\newcommand{\MeV}{\ensuremath{\text{Me\kern -0.1em V}}}
\newcommand{\pt}{\ensuremath{p_{\text{T}}}}

\makeatletter
\g@addto@macro\bfseries\boldmath
\makeatother

\title{A tagger for strange jets based on tracking information using long short-term memory}

\author{Johannes~Erdmann}

\date{\small Technische Universit\"at Dortmund, Dortmund, Germany}

\begin{document}

\maketitle

\begin{abstract}
  An algorithm for the identification of jets that originate from the hadronisation of strange quarks is presented, which complements existing algorithms for the identification of jets that originate from $b$-quarks and $c$-quarks. The algorithm is based on the properties of tracks and uses long short-term memory recurrent neural networks to discriminate between jets from strange quarks and jets from down and up quarks. The performance of the algorithm is compared to a simple benchmark algorithm that uses the transverse-momentum fraction carried by a reconstructed $K_S \rightarrow \pi^+\pi^-$ decay. While the benchmark algorithm is limited to signal efficiencies smaller than 13\%, the proposed algorithm is not limited in efficiency. For signal efficiencies of 30\% and 70\%, background efficiencies of 21\% and 63\% are achieved, indicating the challenge of discriminating strange jets from jets that originate from first-generation quarks.
\end{abstract}

\section{Introduction}
\label{sec:introduction}
\input{introduction}

\section{Simulated samples and jet and track selection}
\label{sec:samples}
\input{samples}

\section{Identification of strange jets}
\label{sec:strangetagging}

\subsection{Using reconstructed $K_S$ mesons}
\label{sec:Kshort}
\input{Kshort}

\subsection{Using long short-term memory}
\label{sec:lstm}
\input{lstm}

\subsection{LSTM using jets with at least one track that does not originate from the primary vertex}
\label{sec:displaced}
\input{displaced}

\subsection{LSTM using all jets}
\label{sec:alltracks}
\input{alltracks}
\clearpage

\section{Results}
\label{sec:results}
\input{results}
\clearpage

\section{Conclusions}
\label{sec:conclusions}
\input{conclusions}

\bibliographystyle{JHEP}
\bibliography{main}

\end{document}

%% file: introduction.tex
Multi-purpose detectors at hadron colliders, such as the ATLAS~\cite{Aad:2008zzm} and CMS~\cite{Chatrchyan:2008aa} experiments at the LHC, are designed to reconstruct and classify the objects that are produced at the interaction point. The identification of jets that originate from the hadronisation of $b$-quarks ($b$-tagging) plays a crucial role for the physics programme at these experiments. Two examples are Higgs-boson~\cite{Aaboud:2018zhk,Sirunyan:2018kst} and top-quark measurements~\cite{Aaboud:2016pbd,Khachatryan:2016kzg} using the $H\rightarrow b\bar{b}$ and $t\rightarrow W^+ b$ decays. Recently, algorithms for the identification of jets that originate from the hadronisation of $c$-quarks ($c$-tagging) have become available~\cite{Aaboud:2018fhh,Sirunyan:2017ezt} and have enabled new possibilities for the analysis of LHC data, notably in the search for the decay $H\rightarrow c\bar{c}$~\cite{Aaboud:2018fhh,Sirunyan:2019qia}.

An algorithm for the identification of jets that originate from the hadronisation of strange quarks (strange tagging) would complement the existing set of algorithms for the identification of the jet flavour, i.e. $b$-tagging, $c$-tagging and algorithms for the discrimination of jets that originate from quarks or gluons (quark--gluon tagging)~\cite{Aad:2014gea,CMS-PAS-JME-13-002}. Such an algorithm would promise to open new opportunities in the analysis of hadron-collider data, such as the direct measurement of the CKM matrix element $V_{ts}$ in the decay $t\rightarrow W^+b$~\cite{Ali:2010xx}, the search for the decay $H\rightarrow s\bar{s}$~\cite{Duarte-Campderros:2018ouv} and the search for new particles that primarily decay to strange quarks.

Algorithms for the identification of the jet flavour strongly rely on machine-learning techniques and in particular deep neural networks~\cite{Guest:2018yhq}. For example, recurrent neural networks have been used for $b$-tagging~\cite{ATL-PHYS-PUB-2017-003}, deep fully-connected neural networks for $c$-tagging~\cite{Aaboud:2018fhh,Sirunyan:2017ezt} and convolutional neural networks for quark--gluon tagging~\cite{Komiske:2016rsd,Aad:2014gea}. Also for the application of identifying high-energetic decays of hadronic resonances, i.e. top quarks, $W$ and $Z$ bosons and Higgs bosons, different machine-learning approaches have been shown to give excellent results~\cite{Guest:2018yhq}.

The key variables for $b$- and $c$-tagging algorithms rely on the lifetime of $B$- and $D$-hadrons, as the decays of these hadrons result in secondary vertices that are spatially separated from the primary vertex (PV). Charged tracks that originate from secondary vertices can be reconstructed with the inner tracking detectors (ID) and show non-zero impact parameters along the direction of the incident hadron beams ($d_z$) and in the plane transverse to it ($d_0$). Analogously, the fragmentation of strange quarks results in the production of strange hadrons. These are in particular kaons and $\Lambda$ baryons, which have a much longer lifetime than $B$- and $D$-hadrons, so that their probability to decay before the first layer of the ID is much lower. However, the lifetimes of $K_S$ mesons and $\Lambda$ baryons are small enough to result in decays within the volume of the ID. The lifetimes of $K_L$ and $K^\pm$ mesons are so large that they mostly do not decay before reaching the calorimeter system.

The goal of a strange-tagging algorithm is to discriminate jets that originate from strange quarks (strange jets) from jets that originate from first-generation quarks ($d$- and $u$-jets), because the discrimination of strange jets from $b$-, $c$- and gluon-jets is possible with the existing $b$-, $c$- and quark--gluon tagging algorithms. While strange hadrons are produced in the fragmentation of strange quarks and their identification was proposed to distinguish strange jets from other jets~\cite{Ali:2010xx,Duarte-Campderros:2018ouv}, the challenge of this classification task is that strange hadrons are also produced frequently in the hadronisation of $d$- and $u$-quarks, however typically with a lower fraction of the jet's transverse momentum (\pt), $x$. The \pt\ fraction of reconstructed $K_S \rightarrow \pi^+\pi^-$ and $\Lambda\rightarrow p\pi^-$ decays\footnote{For simplicity, the $\bar{\Lambda}$ baryon and its decays are not explicitly mentioned in the text, but are also meant when the $\Lambda$ baryon and its decays are discussed.}, $x_K$ and $x_\Lambda$, can hence be used to identify strange jets~\cite{Ali:2010xx}. At detectors that allow for charged-hadron separation with Cherenkov detectors, also the properties of $K^\pm$ mesons can be used~\cite{Abreu:1999cj}, but this possibility often does not exist at multi-purpose detectors, such as ATLAS and CMS.

A strange-tagging algorithm is developed based on the properties of tracks from charged particles, including in particular information about the displacement of the track from the PV. A long short-term memory (LSTM)~\cite{Hochreiter:1997yld} neural network is trained using simulated samples of $s$- and $d$-jets. LSTM networks are a variant of recurrent neural networks (RNNs). Such RNNs are suited for problems that are defined by an ordered series of variable length. RNNs are hence a natural choice for flavour tagging, where each jet is associated with a variable number of tracks and the tracks are ordered according to their properties. LSTM networks have particularly robust training properties for learning features that describe such series. In analogy to time-ordered series, their units do not only include an input and an output gate, but also a ``cell'' that remembers features of earlier time steps and a ``forget gate'' that allows to remove information from earlier time steps from the cell. The performance of the algorithm based on LSTM is compared to the discrimination of the variables $x_K$ and $x_\Lambda$ for reconstructed $K_S\rightarrow\pi^+\pi^-$ and $\Lambda \rightarrow p\pi^-$ decays as a benchmark.

%% file: samples.tex
Simulated samples are generated for the processes $pp\rightarrow s\bar{s}$, $pp\rightarrow d\bar{d}$ and $pp\rightarrow u\bar{u}$ at $\sqrt{s} = 13~\TeV$ with \textsc{MadGraph\_aMC@NLO}~\cite{Alwall:2011uj} at leading order in $\alpha_S$ using the NNPDF2.3LO PDF set~\cite{Ball:2012cx}. Parton showering and hadronisation are simulated with \textsc{Pythia}~8~\cite{Sjostrand:2014zea}. Pile-up collisions are simulated as non-diffractive events with \textsc{Pythia}~8.

A simplified detector is simulated with \textsc{Delphes} version~3.4.1~\cite{deFavereau:2013fsa} using a CMS-like detector as an example for an LHC multi-purpose detector. Jets are built from particle-flow objects with the anti-$k_t$ algorithm~\cite{Cacciari:2008gp} with a distance parameter of 0.5. The efficiency of the ID for the reconstruction of charged hadrons is 70\% (95\%) for $0.1<\pt\leq 1~\GeV$ ($\pt>1~\GeV$) in the central part of the detector ($|\eta|\leq 1.5$, where $\eta$ is the pseudorapidity), and 60\% (85\%) for $0.1<\pt\leq 1~\GeV$ ($\pt>1~\GeV$) for $1.5 < |\eta| \leq 2.5$. The momentum resolution for charged hadrons was modified with respect to the default CMS-like detector simulation that is included in \textsc{Delphes} to correspond closer\footnote{As jets are built from particle-flow objects, including charged hadrons, this modification results in an unrealistically good jet-energy resolution, which is not important for the studies presented here. However, a realistic momentum resolution for charged hadrons is crucial for these studies, which the default CMS card does not provide.} to the resolution in the CMS detector~\cite{Chatrchyan:2014fea}. The resolution is 1\%, 1.5\% and 2\% at $\pt = 0.1~\GeV$ for $|\eta| \leq 0.5$, $0.5 < |\eta| \leq 1.5$ and $1.5 < |\eta| \leq 2.5$, respectively, and it rises to 2\%, 3\% and 7\% at $\pt = 100~\GeV$ for the same regions in pseudorapidity. As in the default CMS detector description in \textsc{Delphes} only the momenta of charged particles have a non-zero resolution, in this analysis additional resolutions are simulated for the flight direction, impact parameters and vertex position of the tracks. For the flight direction, the resolutions of the azimuth angle, $\Phi$, and the cotangent of the polar angle, $\cot\theta$, are taken from Ref.~\cite{Chatrchyan:2014fea}. For $\Phi$ ($\cot\theta$) they amount to 0.008, 0.010 and 0.020 (0.008, 0.015 and 0.05) for $|\eta| \leq 0.5$, $0.5 < |\eta| \leq 1.5$ and $1.5 < |\eta| \leq 2.5$ at $\pt = 0.2~\GeV$, respectively, and they decrease\footnote{The decrease of the resolution is modelled by a heuristic $\sqrt{a^2+b^2/\pt^2}$ formula, with the parameters $a$ and $b$ fit to a set of points taken from the reference publication.} to 0.003, 0.004 and 0.005 (0.006, 0.008 and 0.02) at $\pt = 10~\GeV$ for the same regions in pseudorapidity. The resolutions of $d_0$ and $d_z$ are also taken from Ref.~\cite{Chatrchyan:2014fea}. At $\pt = 0.2~\GeV$ they amount to 0.3~mm, 0.4~mm and 0.7~mm (0.3~mm, 0.6~mm and 2~mm) for $d_0$ ($d_z$) for the three regions in pseudorapidity. The resolutions decrease with increasing $\pt$, resulting in values of 0.02~mm, 0.02~mm and 0.03~mm (0.04~mm, 0.05~mm and 0.1~mm) at $\pt = 10~\GeV$. The resolutions of the $x$-, $y$- and $z$-positions of the tracks' vertices are taken from Refs.~\cite{Chatrchyan:2014fea,CMS-PAS-TRK-10-001} as $\sigma_{x,y}^{(2)} = 0.32$~mm for the $x$- and $y$-positions and $\sigma_{z}^{(2)} = 0.43$~mm for the $z$-position for tracks that originate from vertices with two tracks, and they decrease\footnote{Also here, a similar heuristic function was used to fit a set of points taken from the reference publications: $\sqrt{a^2+b^2/n^2}$, where $n$ is the number of tracks associated to a vertex.} to values of 0.02~mm and 0.025~mm for vertices with 50 tracks. All tracks must fulfil $\pt > 0.1~\GeV$ and $|\eta| < 2.5$.

For the $s\bar{s}$ and $d\bar{d}$ processes, 4,000,000 events are produced each, to be used in the training of the LSTM network. For the $u\bar{u}$ process, 1,000,000 events are produced in order to verify the assumption that the performance for $d$- and $u$-jets is similar. The tracks from pile-up collisions are added to each event, with a number of pile-up interactions according to a Gaussian distribution with a mean of 35 and a width of 15 and assuming a luminous region that follows a Gaussian distribution with a width of 60~mm in $z$-direction. In order to mimic the reconstruction of pile-up vertices, a pile-up vertex is assumed to be reconstructed if it has at least two associated tracks within $5\sigma_{x,y}^{(2)}$ in $x$- and $y$-direction and within $5\sigma_{z}^{(2)}$ in $z$-direction\footnote{The probability for a point to lie within $5\sigma$ in each direction of a three-dimensional Gaussian distribution is 97.7\%.}. Secondary vertices associated with a jet are reconstructed by considering all pairs of tracks matched to the jet within $\Delta R = \sqrt{\Delta\eta^2+\Delta\Phi^2}  = 0.5$. The distance of closest approach of the two tracks must be smaller than $5\sigma_{x,y}^{(2)}$ in $x$- and $y$-direction and smaller than $5\sigma_{z}^{(2)}$ in $z$-direction, and a secondary-vertex candidate is constructed using the centre between the two points of closest approach. Secondary-vertex candidates are rejected if their transverse distance from the PV is either smaller than 4~mm or larger than 450~mm~\cite{Aad:2011hd}. The ambiguity in assigning tracks to secondary-vertex candidates is resolved by removing candidates with tracks that are used multiple times, starting with candidates with large distances from the PV, until no track is used more than once in a candidate.

In each event, the jet with the largest \pt\ is considered, which must fulfil $\pt > 20~\GeV$ as well as $|\eta|<2.5$, so that its axis is within the acceptance of the ID. In order to assign a truth flavor, jets in the $s\bar{s}$ ($d\bar{d}$ or $u\bar{u}$) sample must be geometrically matched to either the $s$-quark ($d$-quark or $u$-quark) or the $\bar{s}$-quark ($\bar{d}$-quark or $\bar{u}$-quark) by requiring $\Delta R < 0.5$. In order to mitigate the effect of pile-up tracks, all tracks are removed that have transverse and longitudinal distance parameters with respect to a reconstructed pile-up vertex smaller than $\sigma_{x,y}^{(2)}$ and $\sigma_{z}^{(2)}$, respectively. This procedure removes pile-up tracks efficiently, adding on average 1.2 pile-up tracks to each jet, but it also removes on average 0.8 tracks that originate from the primary vertex. All remaining tracks that are within $\Delta R = 0.5$ of the jet axis are used. The $\pt$ and $\eta$ spectra of the jets in the $s\bar{s}$ sample and in the $d\bar{d}$ and $u\bar{u}$ samples are very similar. Although the differences are very small---the jets in the $s\bar{s}$ samples have a slightly larger \pt\ on average and tend to be more central than the jets in the other two samples---the jets in the $d\bar{d}$ and $u\bar{u}$ samples are reweighted in \pt\ and $\eta$ to match the spectra in the $s\bar{s}$ sample.

%% file: Kshort.tex
\begin{figure}
  \centering
  \includegraphics[width=0.65\textwidth]{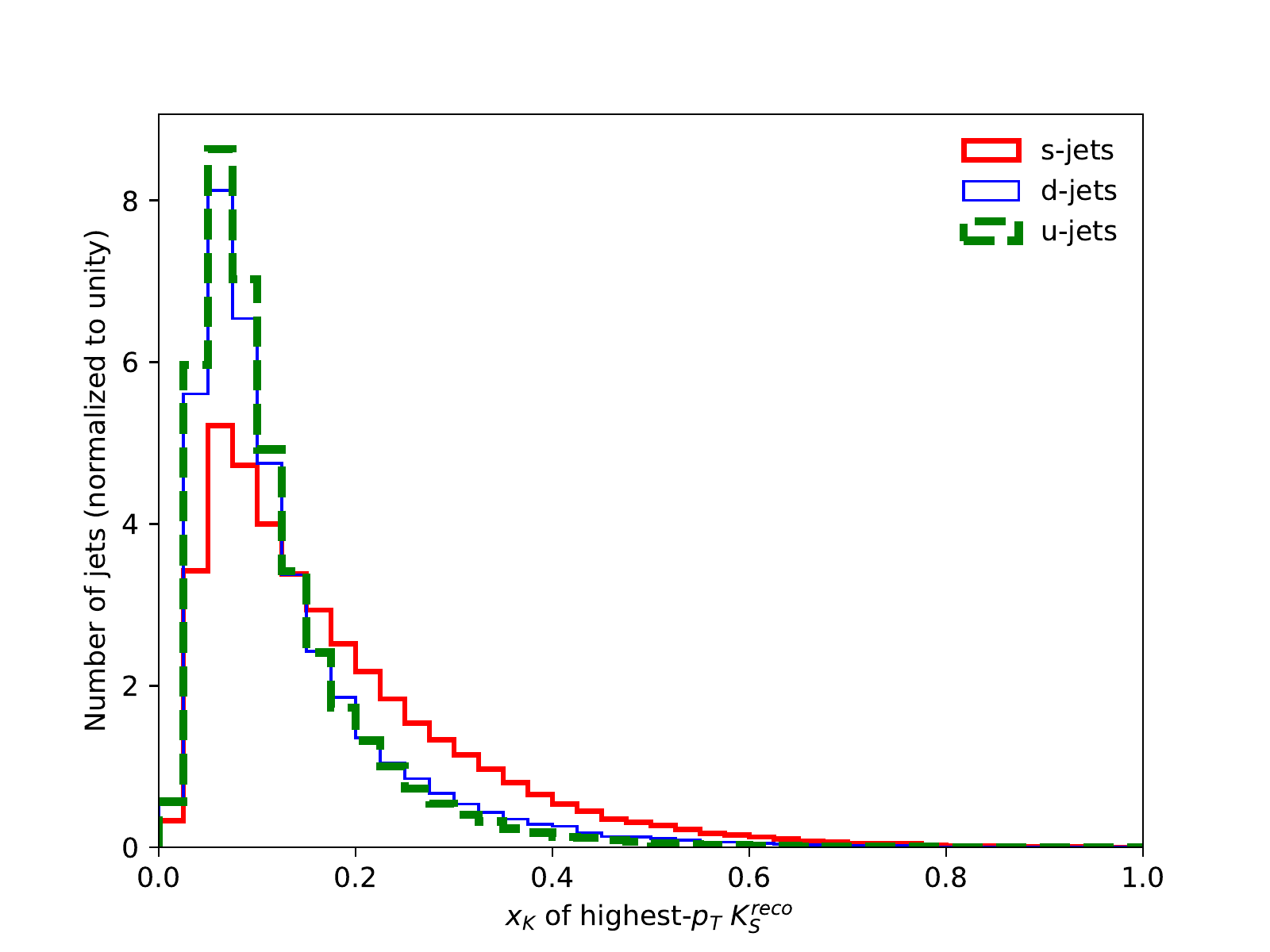} 
  \caption{
    Distribution of $x_K$ for the $K_S$ candidate with the largest \pt\ for $s$-jets, $d$-jets and $u$-jets.
  }
  \label{fig:x_K}
\end{figure}

Decays of $K_S$ mesons to $\pi^+\pi^-$ and $\Lambda$ baryons to $p\pi^-$ are reconstructed by requiring that the sum of the tracks associated with a reconstructed secondary vertex fulfils an invariant-mass criterion. The invariant mass of a $K_S$ candidate must be within the range 480--520~\MeV\ and the invariant mass of a $\Lambda$ candidate must be within 1106--1126~\MeV~\cite{Aad:2019xek}. In the case of the $\Lambda$ baryon, the track with the larger $\pt$ is associated with the proton~\cite{Aad:2019xek}. Approximately 19\% of $s$-jets contain either a reconstructed $K_S$ meson or $\Lambda$ baryon. For $d$- and $u$-jets, this number is 13\%. As shown in Figure~\ref{fig:x_K}, the variable $x_K$ for the reconstructed $K_S$ with the largest \pt\ shows the expected discriminatory power between $s$- and $d$- and $u$-jets.

%% file: lstm.tex
An LSTM recurrent neural network is trained to classify jets as either $s$-jets or $d$-jets using tracking information. Jets from $u$-quarks are not used in the training procedure, but the performance of the neural-network classifier for $d$-jets and $u$-jets is compared after the training and is expected to be similar. The input to the LSTM network is a series of tracks, and the following track properties are used as input to the training process:
\begin{enumerate}
\item the track \pt,
\item the track $\eta$,
\item the signed difference between the track $\eta$ and the jet $\eta$, $\Delta\eta$,
\item the signed difference between the track $\phi$ and the jet $\phi$, $\Delta\phi$,
\item the ratio of the track \pt\ and the jet \pt, $x$,
\item the track momentum perpendicular to the jet axis, $\pt^\mathrm{rel}$,
\item the track momentum in the direction of the jet axis, $p_z^\mathrm{rel}$,
\item the transverse impact parameter of the track, $d_0$,
\item the longitudinal impact parameter of the track, $d_z$,
\item the distance in $x$-direction, $\Delta x$, between the PV and either the associated secondary vertex (if the track is associated to a secondary vertex) or the position of the track's first hit in the ID (otherwise),
\item $\Delta y$, defined analogously to $\Delta x$,
\item $\Delta z$, defined analogously to $\Delta x$.
\end{enumerate}
In addition, the jet \pt\ and $\eta$ are used as input to the training process separately, i.e. not as part of the track series, as described below.

In the ID, silicon layers are assumed to be present at radii\footnote{Although a CMS-like detector is used in \textsc{Delphes}, the layout of the ID silicon layers is inspired by the layout of the ATLAS detector, as also the reconstruction of $K_S$ mesons and $\Lambda$ baryons follows the strategy of an ATLAS analysis. The details of the detector simulation are expected to have a small impact on the results of this work.}~\cite{Aad:2008zzm} $R = 50~\mathrm{mm}$, 90~mm, 120~mm, 300~mm, 370~mm, 440~mm and 510~mm. The positions of a track's first hit in the ID is hence determined based on this layout. Tracks that result from decays in flight outside of the ID volume are not considered. Although a hit efficiency of 100\% is assumed, silicon tracking detectors indeed achieve values close to 100\%, so that using a realistic hit efficiency is expected to have a negligible impact on the results.

The information that is used for each track is redundant by construction in order to facilitate the training of the network: track momenta and angles are provided as well as relative momenta with respect to the jet \pt\ and angular differences with respect to the jet axis; variables that describe the presence of tracks that do not originate from the PV are provided in terms of impact parameters and in terms of the coordinates of either the associated secondary vertex or of the position of the first hit in the ID.

Important information for the classification task is expected to be connected to the presence of displaced decays of strange hadrons. Hence, the track series is ordered by $R = \sqrt{\left(\Delta x)\right)^2 + \left(\Delta y)\right)^2}$, i.e. the transverse distance from the PV of either the associated secondary vertex or of the first hit in the ID. If $R$ is the same for several tracks, the tracks are ordered by their \pt. Consequently, tracks that are associated with the same secondary vertex have adjacent positions in the track series, so that the network is expected to learn from the presence of displaced decays.

For the training, all tracks are used that are associated with a jet. The maximum number of tracks associated with one jet in the simulated samples is 36. All input features are scaled with \textsc{scikit-learn}'s~\cite{Pedregosa:2012toh} RobustScaler, which transforms each feature by subtracting its median and dividing it by the interquartile range.

The first layer of the network is an LSTM layer with a certain number of units. The output of the LSTM layer is equal to the number of its units and is used as input to a fully-connected network with a certain number of layers (``dense layers''), together with the jet \pt\ and the jet $\eta$. These jet variables are used as additional input features, so that the network can learn from correlations of the track series with the jet kinematics. The two-dimensional reweighting of the $d$-jet sample to the jet \pt\ and $\eta$ spectrum of the $s$-jet sample prevents the network to discriminate signal and background jets based on their kinematics alone. The number of nodes in each layer of the fully-connected network is given by half of the number of nodes in the previous dense layer. The number of units in the LSTM layer, the number of dense layers and the number of nodes in the first dense layer are hyperparameters of the network structure. The activation functions are the hyperbolic tangent and the sigmoid function as cell and gate activations in the LSTM layers, respectively, the sigmoid function in the last fully-connected layer and the rectified linear unit in all other fully-connected layers.

The network is implemented with \textsc{Keras}~\cite{chollet2015keras} using \textsc{TensorFlow}~\cite{tensorflow2015-whitepaper} as the back end. The simulated samples are split into a training sample (60\%), a validation sample (20\%) for the evaluation of the performance when the hyperparameters of the network are changed and an independent test sample (20\%) for the evaluation of the performance. All of these samples consist of 50\% $s$-jets and 50\% $d$-jets.

%% file: displaced.tex
As the presence of tracks that do not originate from the PV is expected to provide main discrimination power between signal and background jets, as a first step an LSTM network is trained using only jets with at least one track with $|d_0| > 1~\mathrm{mm}$. Out of the sample of $s$-jets, 74\% of the jets fulfil this criterion. The corresponding number in the $d$-jets sample is 71\%. As this additional requirement slightly changes the \pt\ and $\eta$ spectra of the jets, an alternative reweighting of the $d$-jet sample to the $s$-jet sample is performed in order to prevent the network to learn from (slight) differences in the jet kinematics. This alternative reweighting is only applied for the training procedure. For the comparison of the different methods in Section~\ref{sec:results}, the nominal reweighting as introduced in Section~\ref{sec:samples} is used in order to allow for a consistent comparison of the different methods.

The network is trained using the Adam optimizer~\cite{Kingma:2014vow}. The batch size and the learning rate are hyperparameters of the Adam optimiser that define the sample size over which gradients are calculated during stochastic gradient descent and the step size in the optimisation, respectively. A batch size of 4000 jets and a learning rate of $10^{-5}$ are chosen, which are found to provide stable training results. Binary cross entropy is used as the loss function for the training. The training procedure is stopped if over the course of 10 iterations over the entire training sample (epochs), the value of the loss function that is evaluated on the validation sample does not improve. The epoch with the minimal loss in the validation sample is chosen as the best training.

\begin{figure}[p]
  \centering
  \subfloat[]{\includegraphics[width=0.7\textwidth]{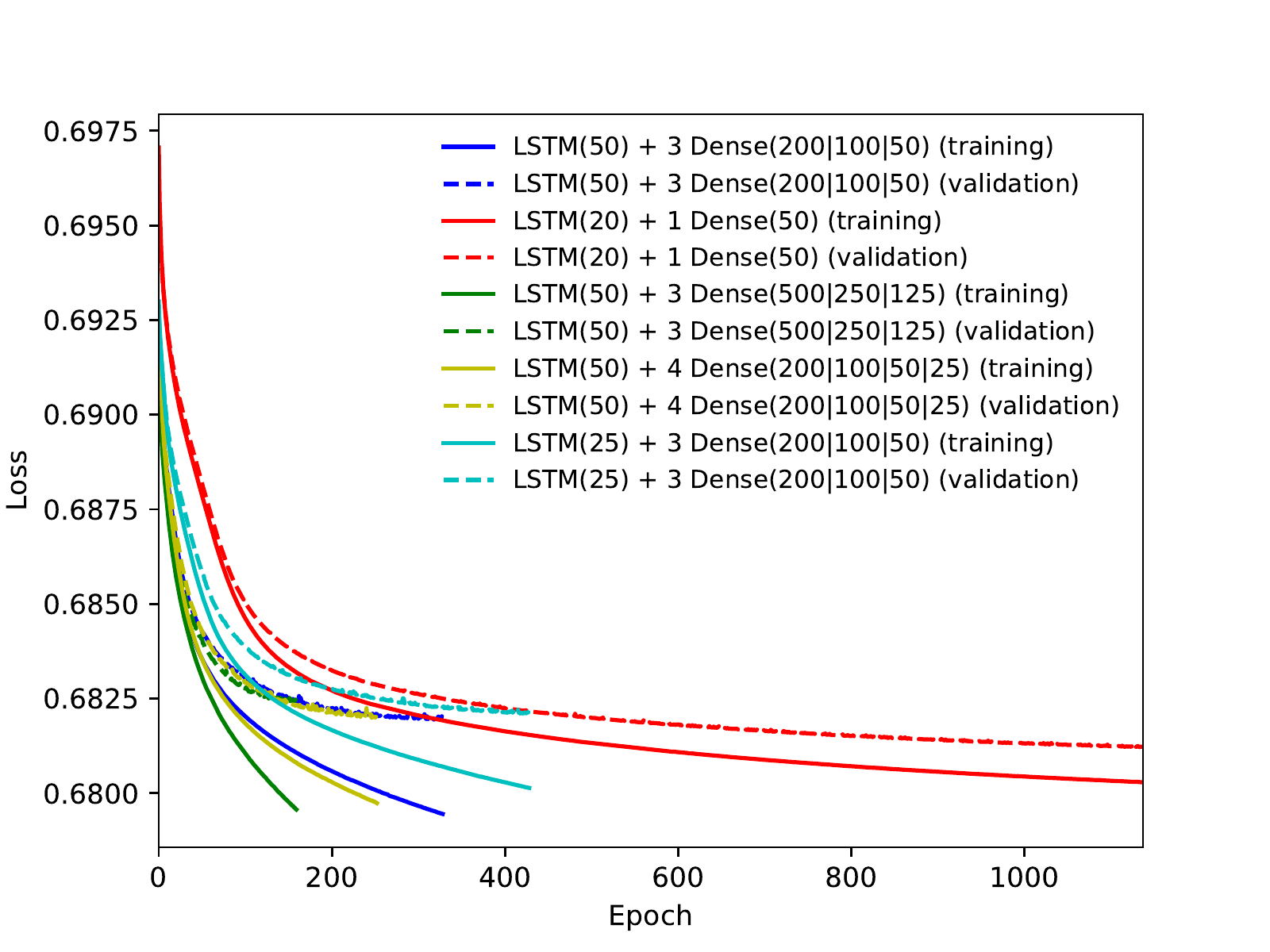}}\\
  \subfloat[]{\includegraphics[width=0.7\textwidth]{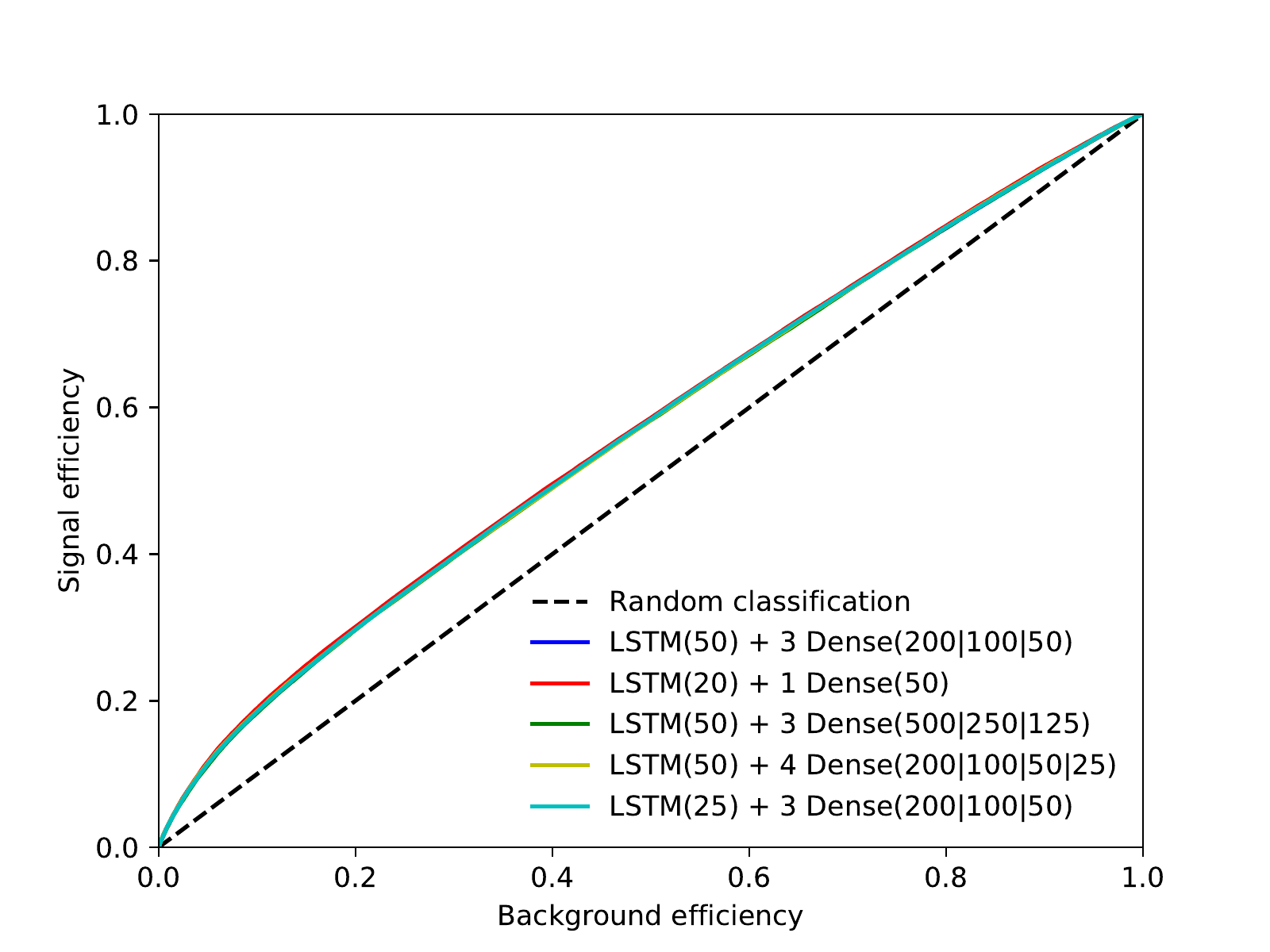}}
  \caption{
    (a) Value of the loss function as a function of the training epoch for different neural-network configurations when the network is trained with jets that have at least one track that does not originate from the primary vertex ($|d_0| > 1~\mathrm{mm}$). The value of the loss function is evaluated on the training sample (solid) and on the validation sample (dashed). (b) ROC curve for different neural-network configurations evaluated on the test sample.
  }
  \label{fig:lossroc1}
\end{figure}

As nominal network structure an LSTM layer with 50 units is chosen that is connected to 3 dense layers with 200, 100 and 50 nodes, each. The other network structures are variations of the nominal structure with a larger or smaller number of LSTM units, a larger or smaller number of dense layers and a larger or smaller number of nodes in the first dense layer. The value of the loss function is shown in Figure~\ref{fig:lossroc1}(a) as a function of the training epoch for these configurations for the training sample and for the validation sample. The values of the loss function in the validation sample converge towards a constant value during the training process. The corresponding receiver operating characteristic (ROC) curves are shown in Figure~\ref{fig:lossroc1}(b), where the efficiency for the identification of $s$-jets (signal efficiency) is shown as a function of the efficiency for $d$-jets (background efficiency). While the performance is very similar for the different network structures, the best training is reached after a different number of epochs. While for the network structures with fewer parameters a large number of epochs need to be trained, the training stops earlier for networks with a larger number of parameters. As no large differences between the different choices for the hyperparameters are observed, the nominal network structure is chosen for further studies.

\begin{figure}
  \centering
  \includegraphics[width=0.7\textwidth]{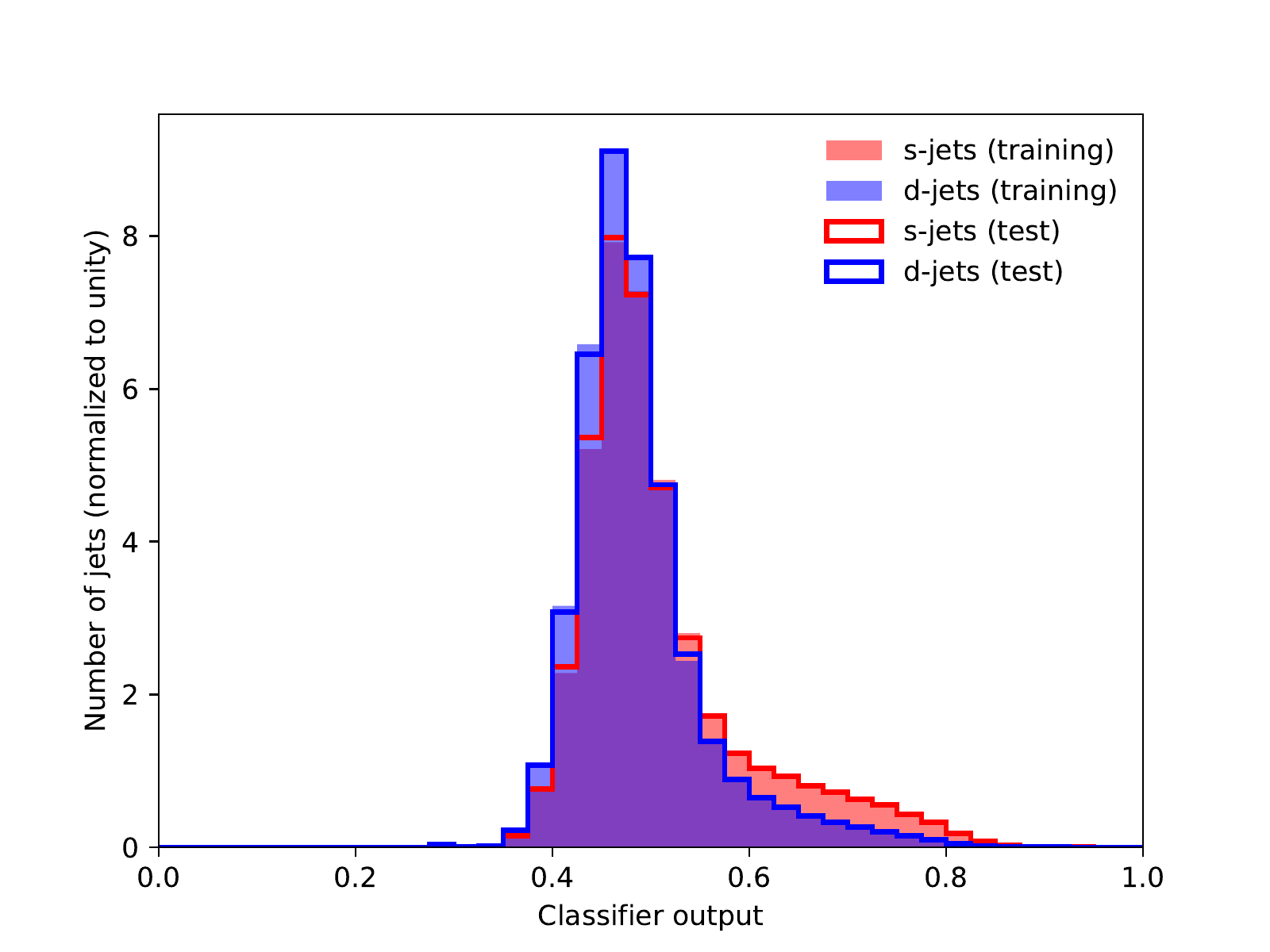}
  \caption{
    Distribution of the classifier output for $s$- and $d$-jets for the network with 50 LSTM units and 3 fully-connected layers with 200, 100 and 50 nodes evaluated on the training and on the test sample, trained with jets that have at least one track that does not originate from the primary vertex ($|d_0| > 1~\mathrm{mm}$).
  }
  \label{fig:output1}
\end{figure}

The distribution of the classifier output for $s$- and $d$-jets is shown in Figure~\ref{fig:output1} for the nominal network, comparing the distributions on the training and on the test sample. Signal and background are separated in both samples and the distributions are very similar when evaluated on the training and on the test sample, showing no indication of overtraining.

%% file: alltracks.tex
\begin{figure}[p]
  \centering
  \subfloat[]{\includegraphics[width=0.7\textwidth]{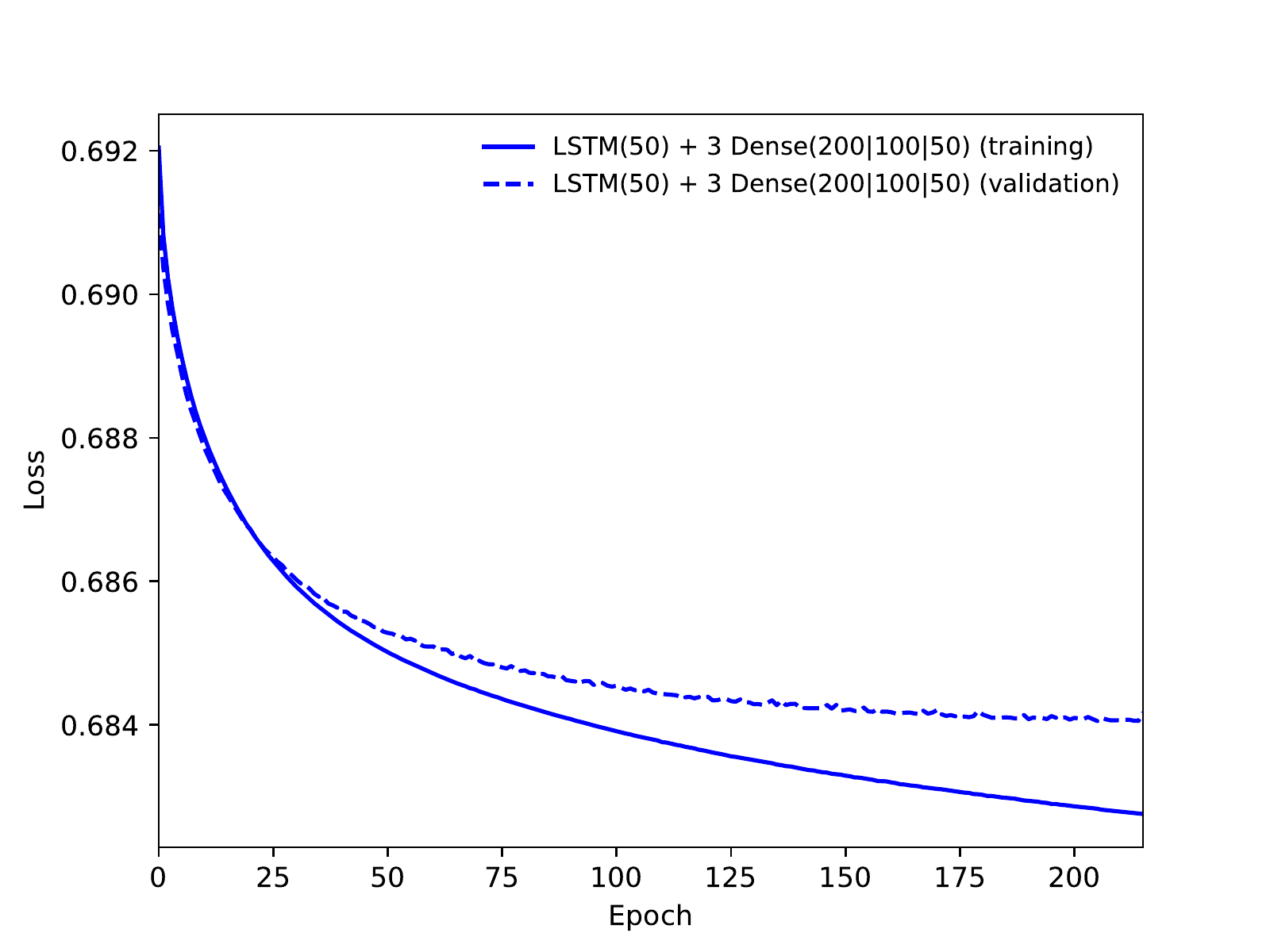}}\\
  \subfloat[]{\includegraphics[width=0.7\textwidth]{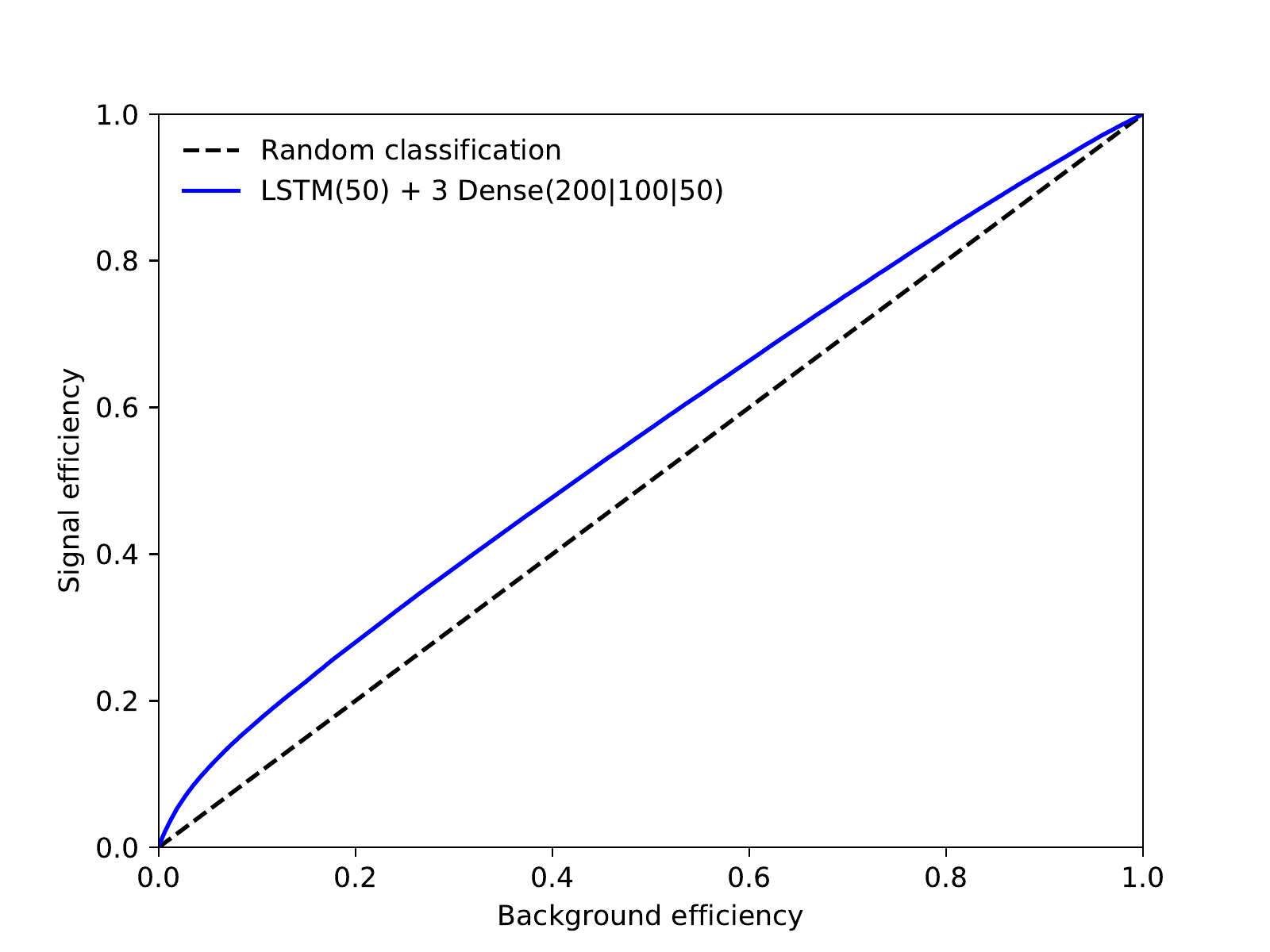}}
  \caption{
    (a) Value of the loss function as a function of the training epoch for the network with 50 LSTM units and 3 fully-connected layers with 200, 100 and 50 nodes trained with all jets. The value of the loss function is evaluated on the training sample (solid) and on the validation sample (dashed). (b) ROC curve evaluated on the test sample.
  }
  \label{fig:lossroc2}
\end{figure}

In a second step, a network is trained with the nominal network structure as introduced in Section~\ref{sec:displaced}, i.e. 50 LSTM units, 3 dense layers with 200, 100 and 50 nodes, each, now using all jets. The value of the loss function is shown in Figure~\ref{fig:lossroc2}(a) as a function of the training epoch evaluated on the training and validation sample, and the ROC curve is shown in Figure~\ref{fig:lossroc2}(b). The training stops after 216 epochs. This network is not limited in the reach in signal efficiency, unlike the strange-tagging algorithms discussed in Sections~\ref{sec:Kshort} and~\ref{sec:displaced}, i.e. it is able to discriminate $s$- and $d$-jets using the pattern of associated tracks even if no tracks are found that do not originate from the PV.

\begin{figure}[p]
  \centering
  \subfloat[]{\includegraphics[width=0.7\textwidth]{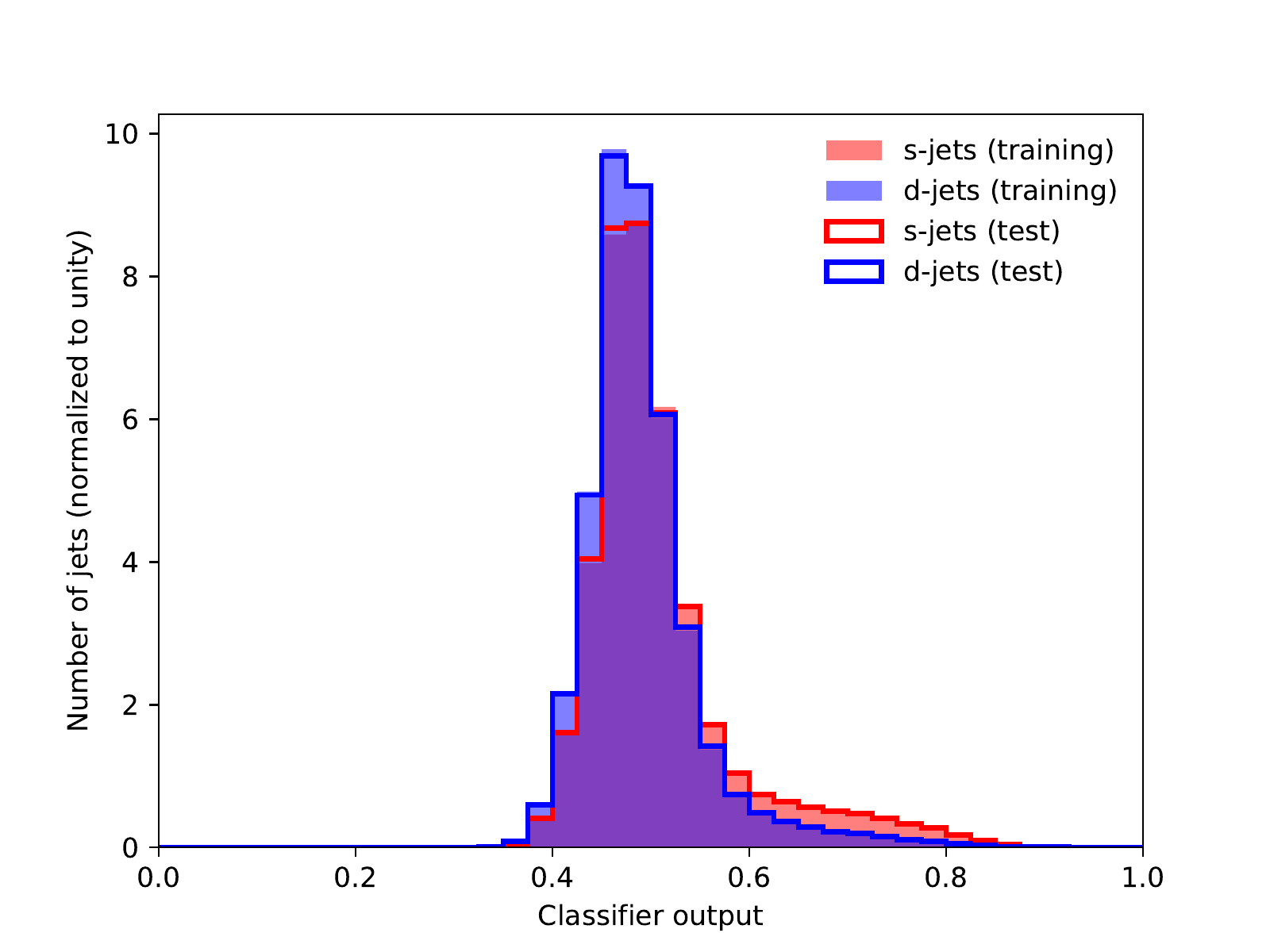}}\\
  \subfloat[]{\includegraphics[width=0.7\textwidth]{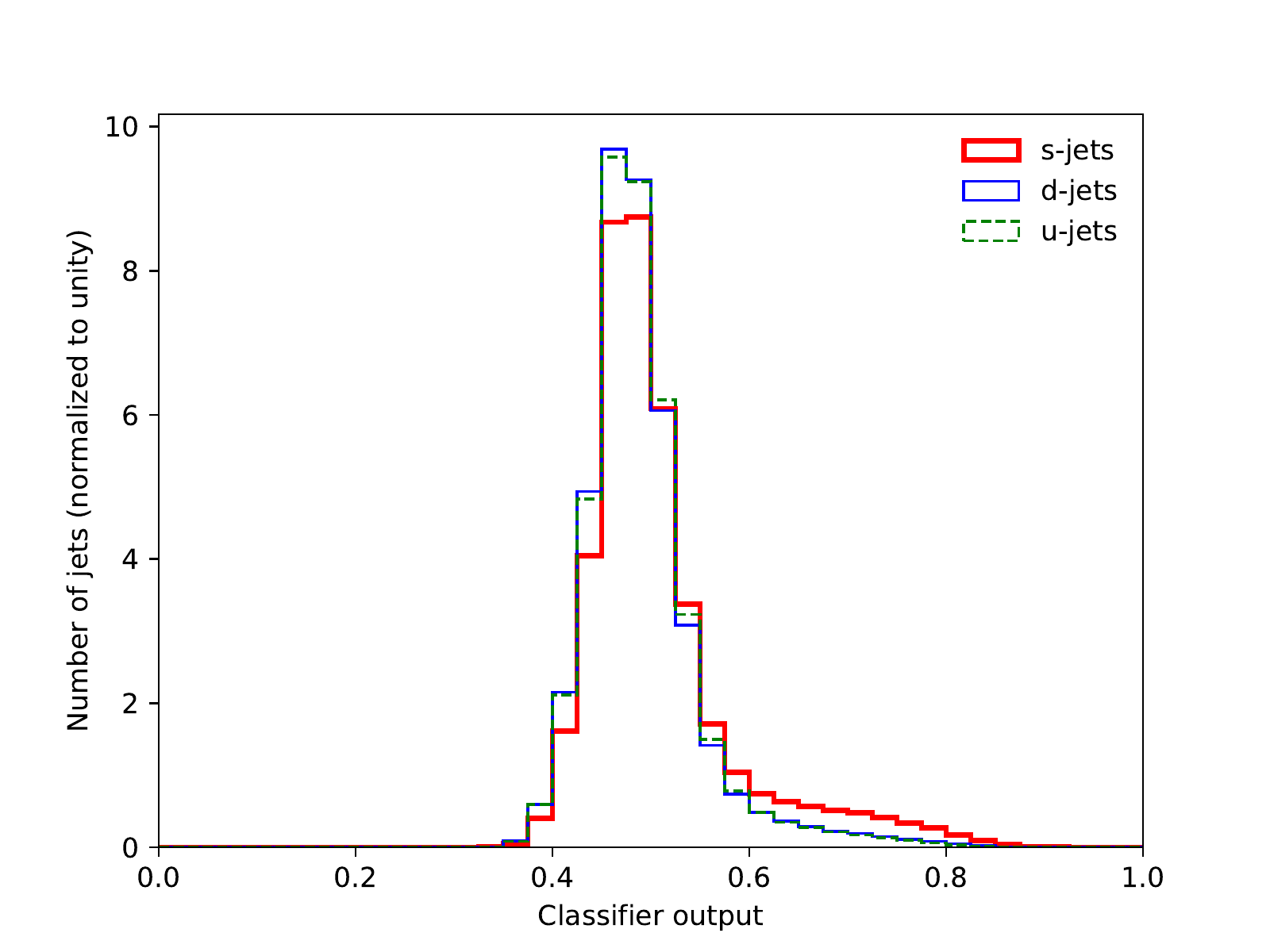}}
  \caption{
    Distribution of the classifier output for the network with 50 LSTM units and 3 fully-connected layers with 200, 100 and 50 nodes, trained with all jets, (a) evaluated on the training and on the test sample for $s$- and $d$-jets and (b) evaluated on the test sample for $s$-, $d$- and $u$-jets.
  }
  \label{fig:output2}
\end{figure}

As in Section~\ref{sec:displaced}, the distributions of the classifier output evaluated on the training and on the test sample do not show an indication of overtraining (Figure~\ref{fig:output2}(a)). The classifier output on the test sample is compared for $s$-, $d$- and $u$-jets in Figure~\ref{fig:output2}(b). While the separation between $s$-jets and $d$- and $u$-jets is clearly visible, the distributions for $d$- and $u$-jets are similar, although only $d$-jets were used as background in the training.

\begin{figure}
  \centering
  \includegraphics[width=0.7\textwidth]{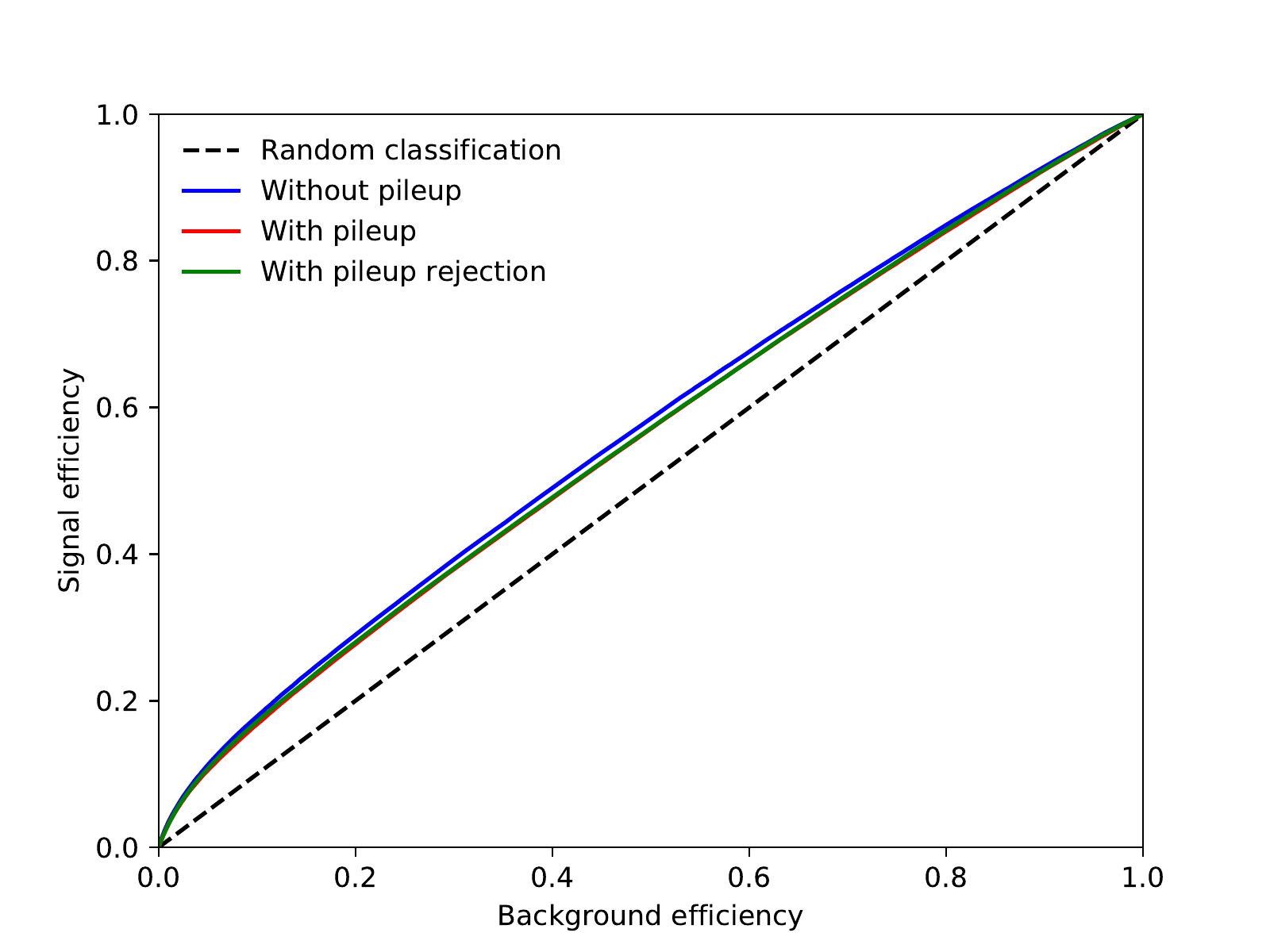}
  \caption{
    ROC curves evaluated on test sample for network trainings (50 LSTM units, 3 fully-connected layers with 200, 100 and 50 nodes) with different pile-up configurations: nominal training, i.e. including pile-up tracks as well as the rejection of pile-up tracks (``With pileup rejection''), a training including pile-up tracks but no rejection of such tracks (``With pileup'') and a training that does not include pile-up tracks (``Without pileup'').  The curve ``With pileup'' is almost entirely hidden below the curve ``With pileup rejection''. 
  }
  \label{fig:pileup}
\end{figure}

The effect of pile-up on the classifier is studied by comparing the nominal setup, which includes pile-up tracks as well as a criterion for the rejection of such tracks, to alternative setups that either do not include pile-up effects at all or for which pile-up tracks are present but no rejection criterion is applied. For each of these setups, an LSTM network is trained (50 LSTM units, 3 fully-connected layers with 200, 100 and 50 nodes) and the ROC curves, evaluated on the respective test samples, are compared in Figure~\ref{fig:pileup}. While it is visible that the inclusion of pile-up deteriorates the performance, its effect is not dramatic. It is interesting, however, that the performance for the cases that include pile-up but differ by the pile-up rejection criterion is small. The gain from rejecting pile-up tracks may be balanced by losing tracks that originated from the hard-scattering vertex. Although the competitive performance of the network trained with all pile-up tracks indicates that the LSTM network is able to learn information from the pile-up pattern, such an approach may be subject to large systematic uncertainties in the pile-up modelling and a training including pile-up rejection may be more useful at an experiment.

%% file: results.tex
\begin{figure}[h!]
  \centering
  \includegraphics[width=0.7\textwidth]{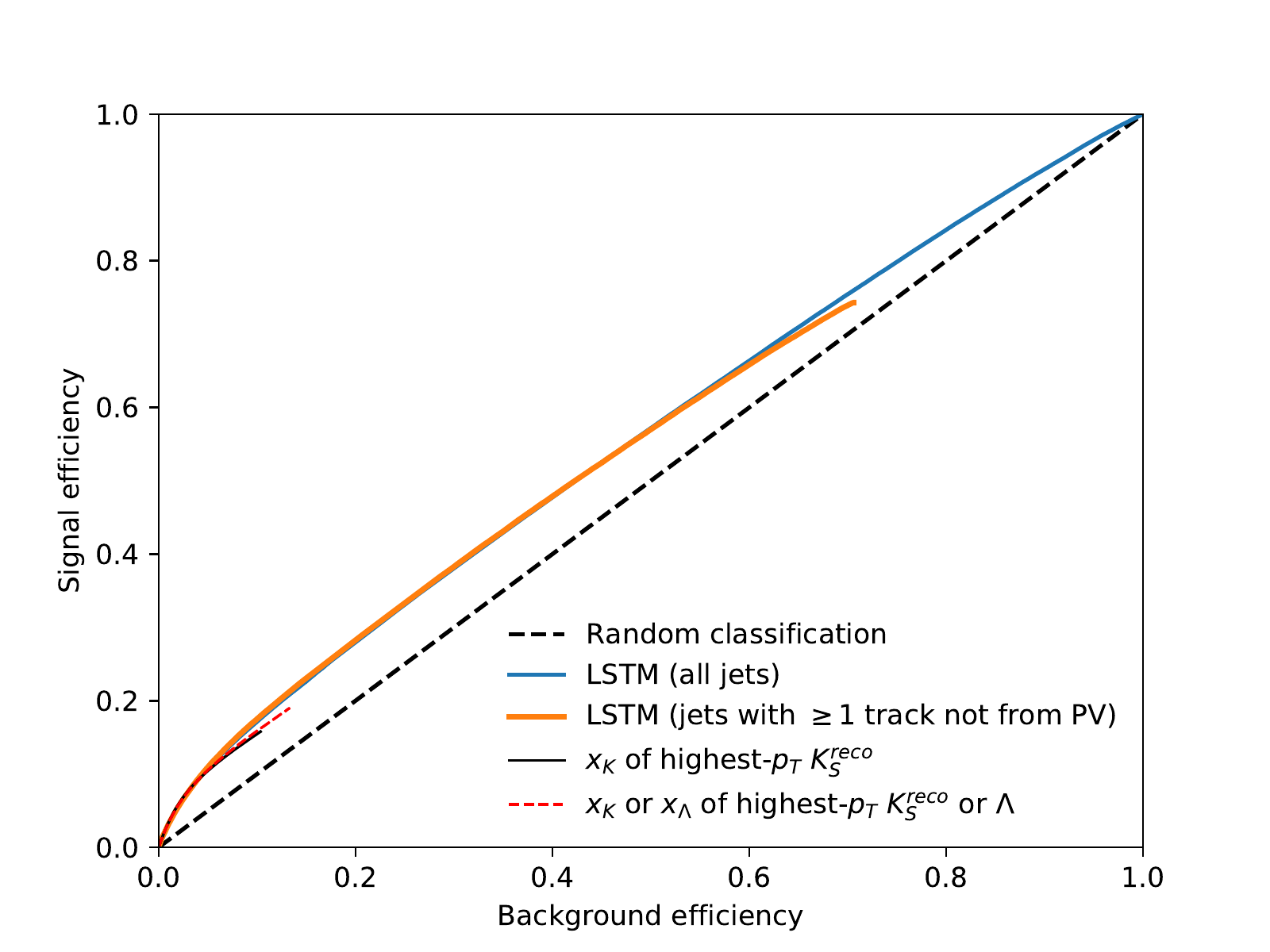}
  \caption{
    ROC curves evaluated on the test sample for the networks with 50 LSTM units and 3 fully-connected layers with 200, 100 and 50 nodes trained with all jets (blue solid line), trained only with jets that have at least one track that does not originate from the primary vertex (orange solid line), for $x_K$ of the $K_S$ candidate with the largest \pt\ in the jet (thin black line) and for using also $x_{\mathrm{\Lambda}}$ of the $\Lambda$ candidate with the largest \pt\ if no $K_S$ candidate is found (thin red dashed line). The latter three ROC curves end at the efficiencies corresponding to the presence of at least one track that does not originate from the primary vertex ($|d_0| > 1~\mathrm{mm}$), the presence of at least one reconstructed $K_S\rightarrow\pi^+\pi^-$ decay, and the presence of at least one reconstructed $K_S\rightarrow\pi^+\pi^-$ or $\Lambda\rightarrow p\pi^-$ decay.
    }
  \label{fig:allrocs}
\end{figure}

Strange tagging based on $x_K$ of the highest-\pt\ reconstructed $K_S$ and based on the two LSTM networks introduced in Section~\ref{sec:strangetagging} are compared in terms of their performance in Figure~\ref{fig:allrocs}, which shows ROC curves for these three methods, as well as an extension of the $x_K$ method, where $x_{\Lambda}$ of the highest-\pt\ reconstructed $\Lambda$ baryon is used if no reconstructed $K_S$ is found in the event. The ROC curves for $x_K$, $x_K$\&$x_{\Lambda}$ and for the LSTM network trained with jets that have at least one track not from the PV are limited in their efficiencies, as discussed in Sections~\ref{sec:Kshort} and~\ref{sec:displaced}. Within their efficiency reach, however, the ROC curves of the different methods show a similar performance. The advantage of the LSTM networks does hence not lie in an improved signal--background separation, but in a larger range of accessible signal efficiencies, which may be beneficial for the application of strange tagging in data analysis.

\begin{figure}
  \centering
  \subfloat[]{\includegraphics[width=0.49\textwidth]{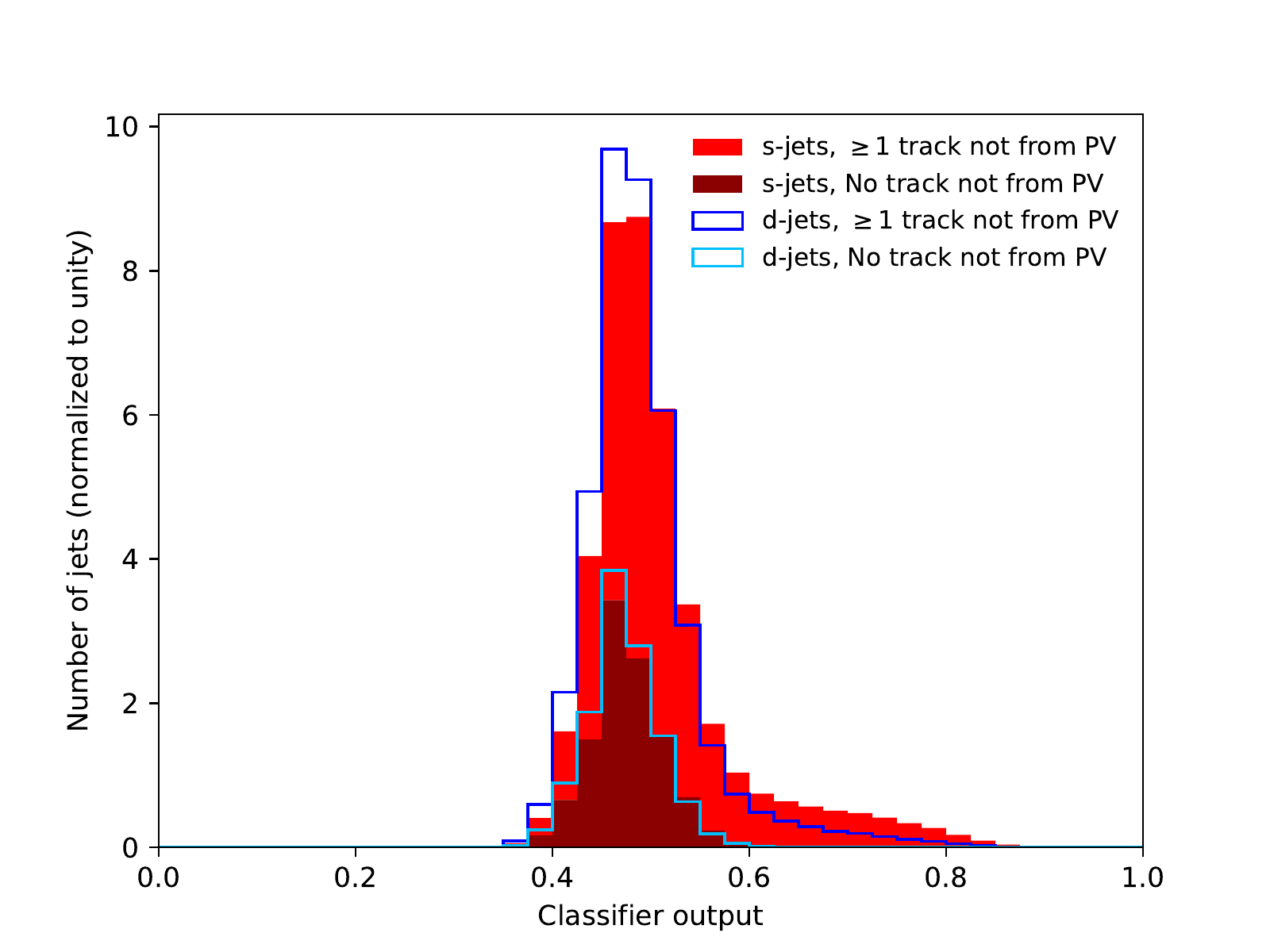}}
  \subfloat[]{\includegraphics[width=0.49\textwidth]{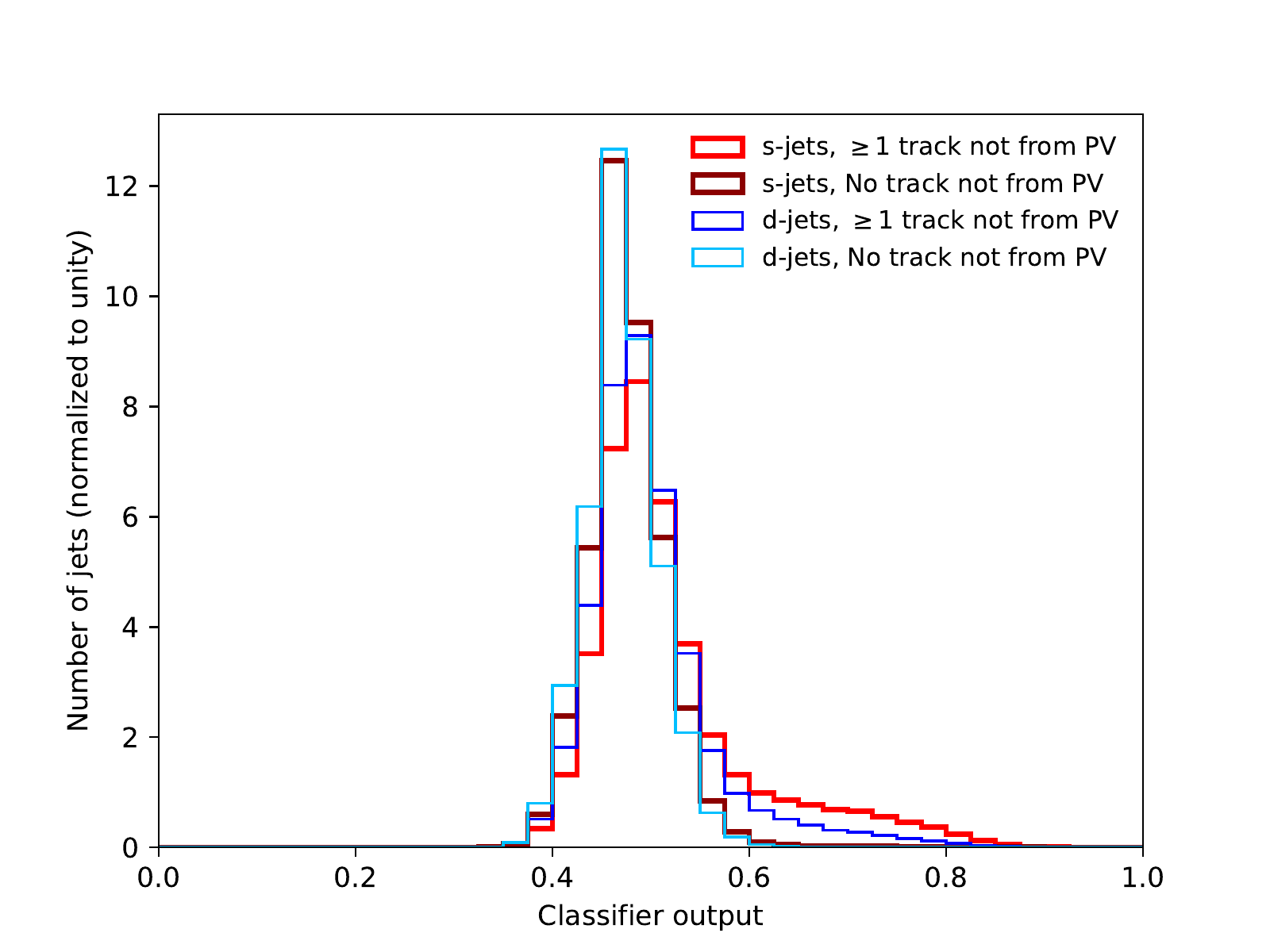}}\\
  \caption{
    Distribution of the classifier output evaluated on the test sample for $s$- (red) and $d$-jets (blue) for the network with 50 LSTM units and 3 fully-connected layers with 200, 100 and 50 nodes trained with all jets. The distributions are split by requiring either at least one track that does not originate from the primary vertex ($|d_0|>1~\mathrm{mm}$) or no such track. In figure (a), the split distributions are stacked on top of each other for $s$- and $d$-jets separately. In figure (b), the split distributions are shown separately and are each normalised to unity.
  }
  \label{fig:inv1}
\end{figure}

\begin{figure}
  \centering
  \subfloat[]{\includegraphics[width=0.49\textwidth]{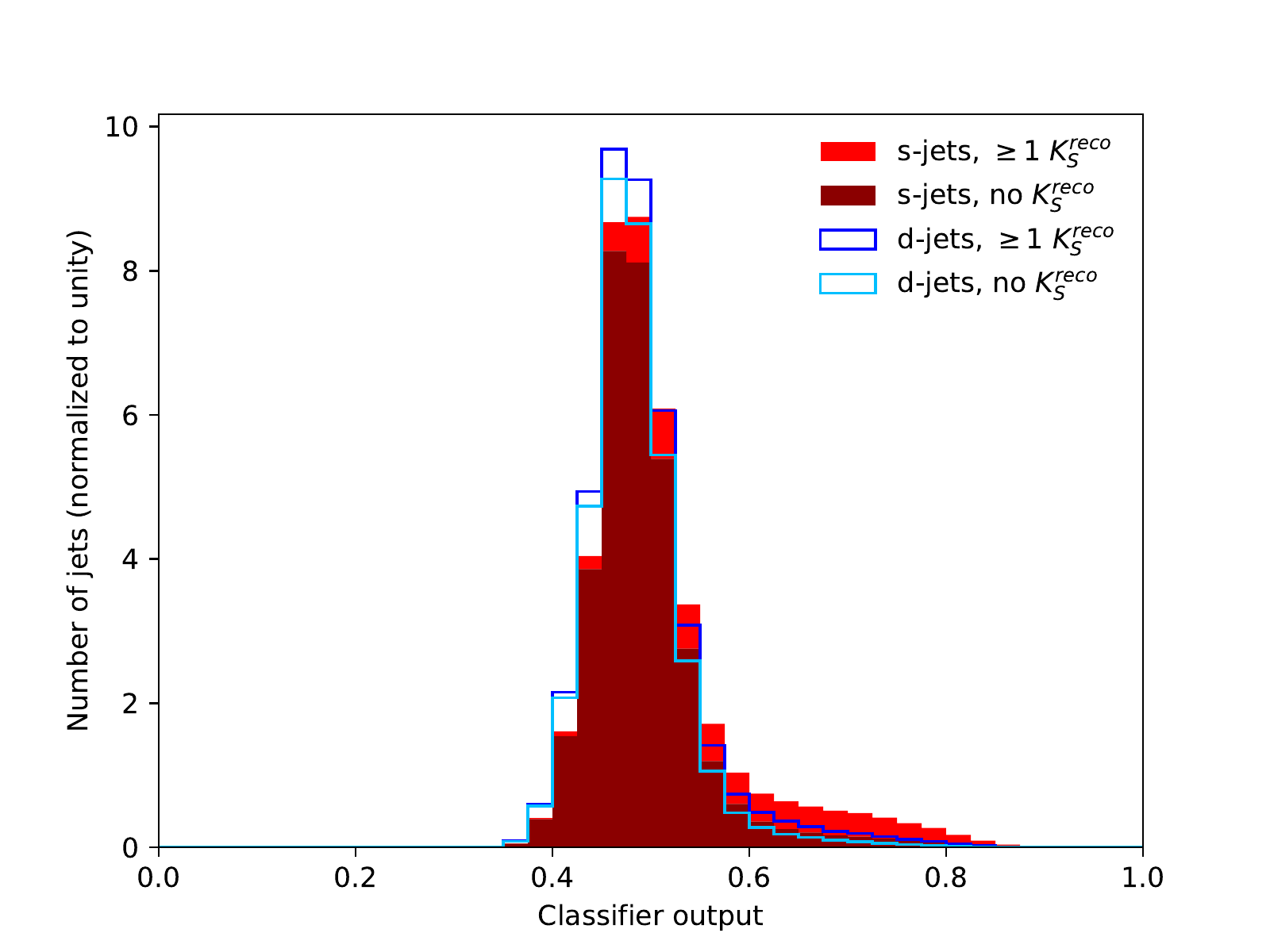}}
  \subfloat[]{\includegraphics[width=0.49\textwidth]{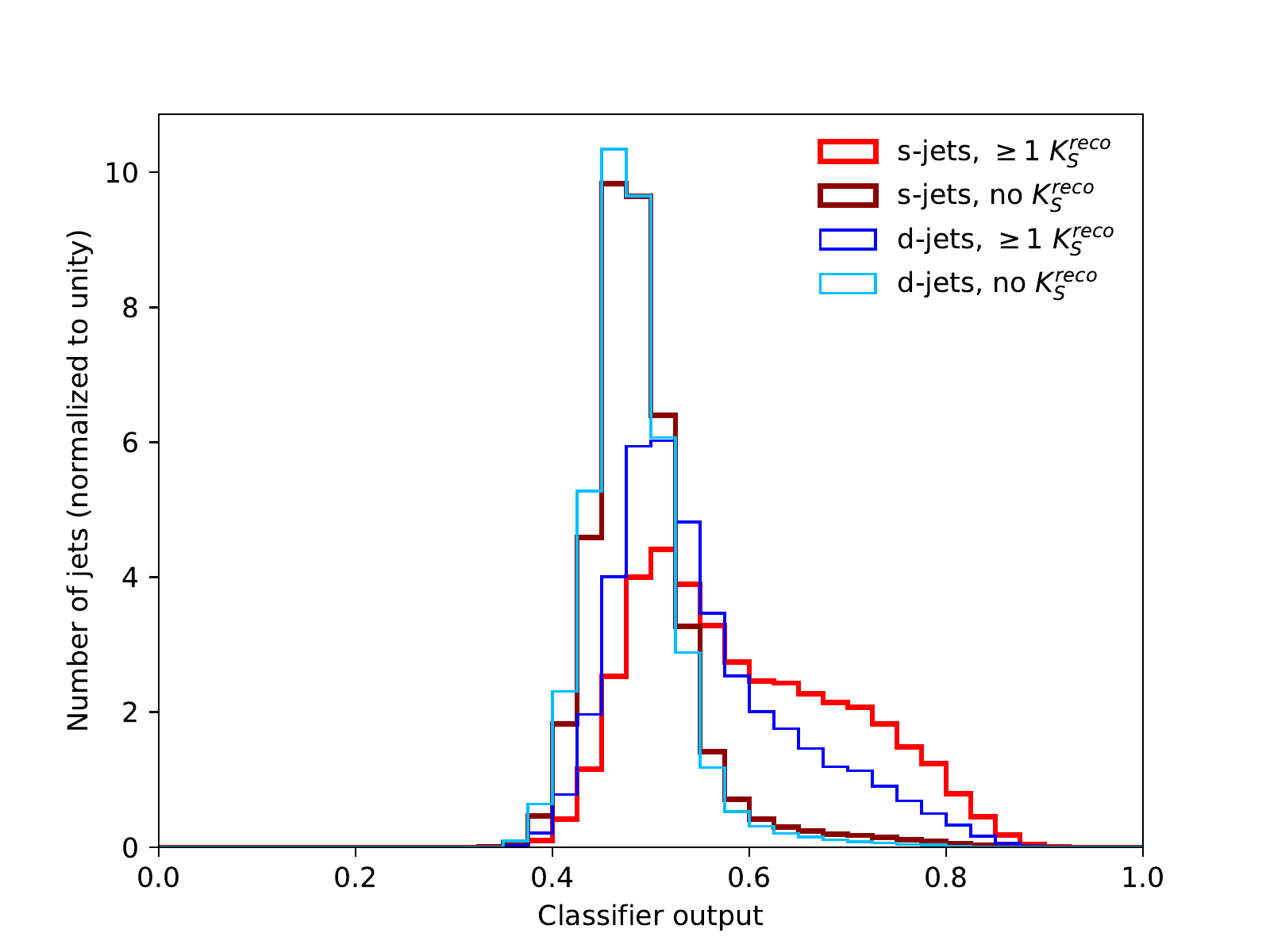}}\\
  \subfloat[]{\includegraphics[width=0.49\textwidth]{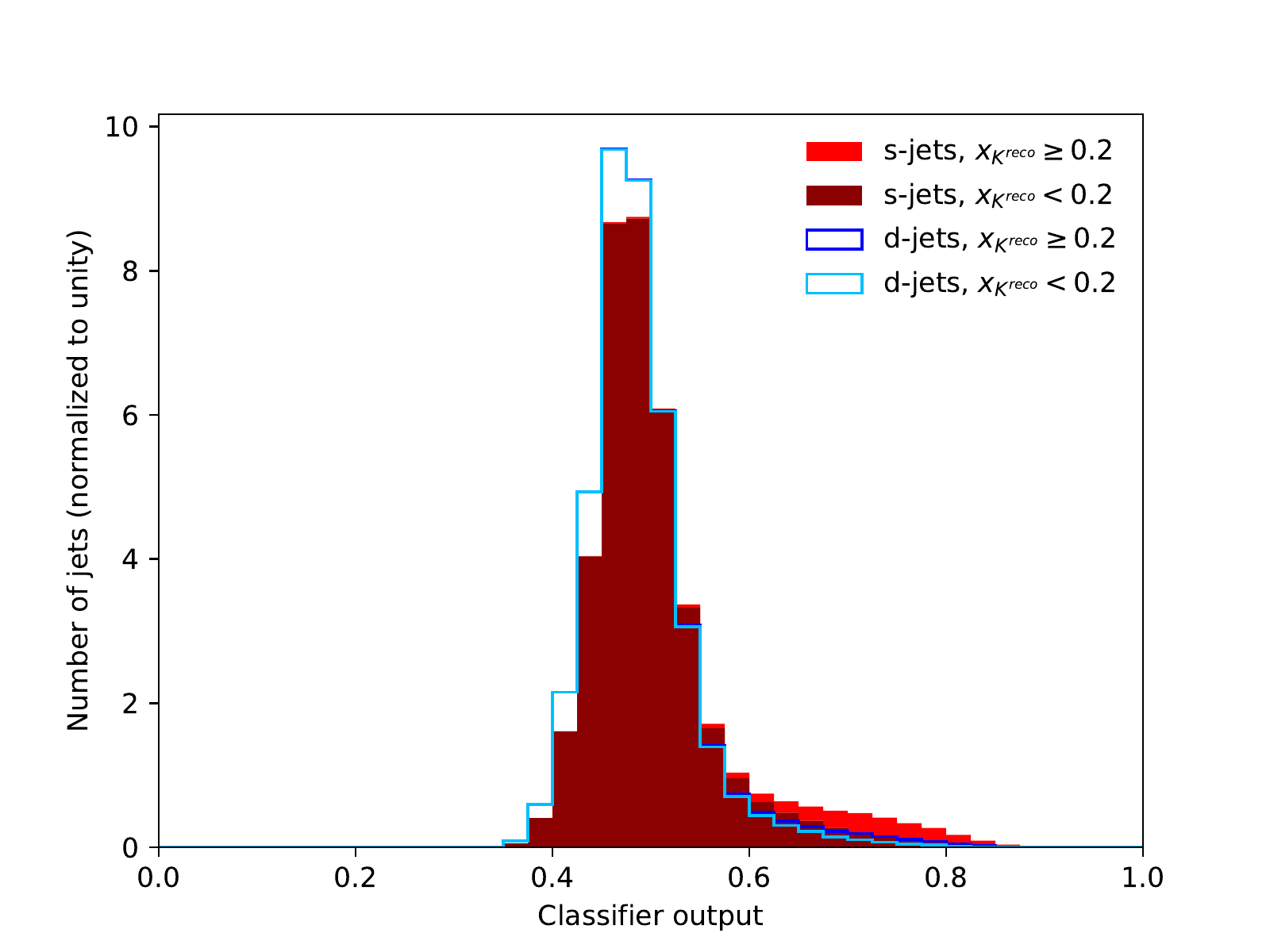}}
  \subfloat[]{\includegraphics[width=0.49\textwidth]{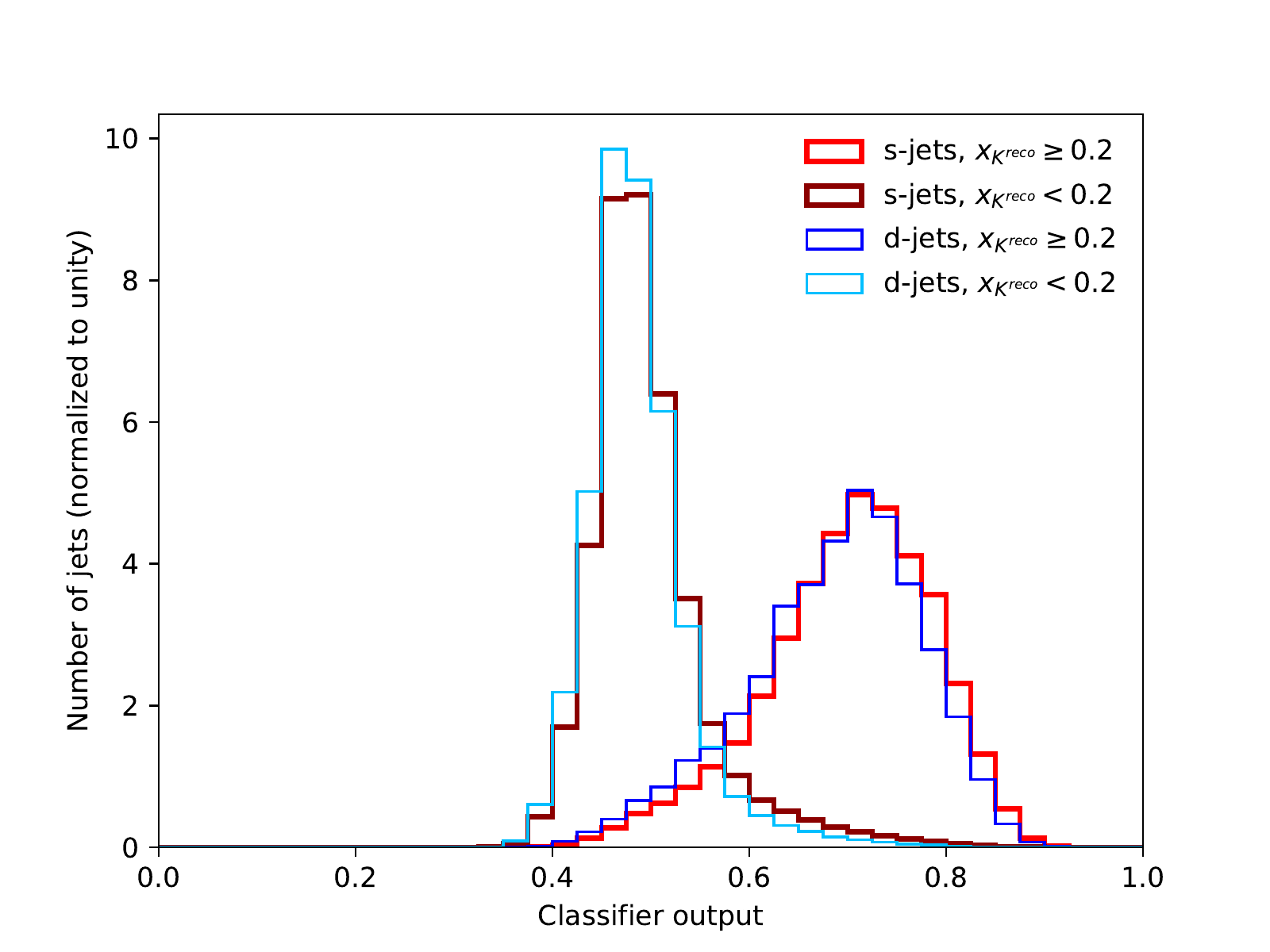}}\\
  \caption{
    Distribution of the classifier output evaluated on the test sample for $s$- (red) and $d$-jets (blue) for the network with 50 LSTM units and 3 fully-connected layers with 200, 100 and 50 nodes trained with all jets. The distributions are split (a,b) by requiring at least one reconstructed $K_S$ candidate or no such candidate or (c,d) by requiring $x_K$ of such a candidate to be either larger or smaller than 0.2 (jets without reconstructed $K_S$ candidates are counted with an $x_K$ of 0). In figures (a) and (c), the split distributions are stacked on top of each other for $s$- and $d$-jets separately. In figures (b) and (d), the split distributions are shown separately and are each normalised to unity.
  }
  \label{fig:inv2}
\end{figure}

\begin{figure}
  \centering
  \subfloat[]{\includegraphics[width=0.49\textwidth]{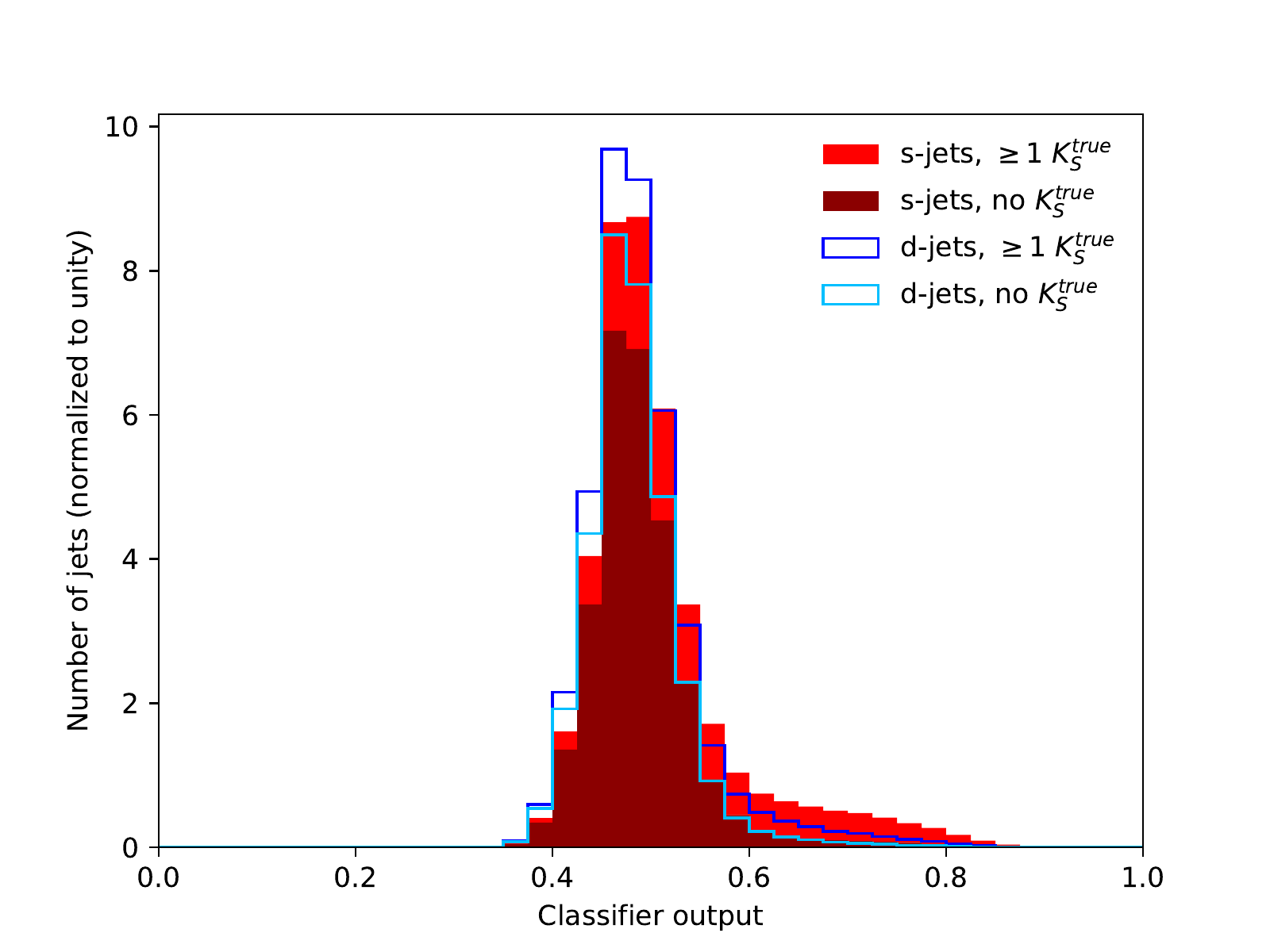}}
  \subfloat[]{\includegraphics[width=0.49\textwidth]{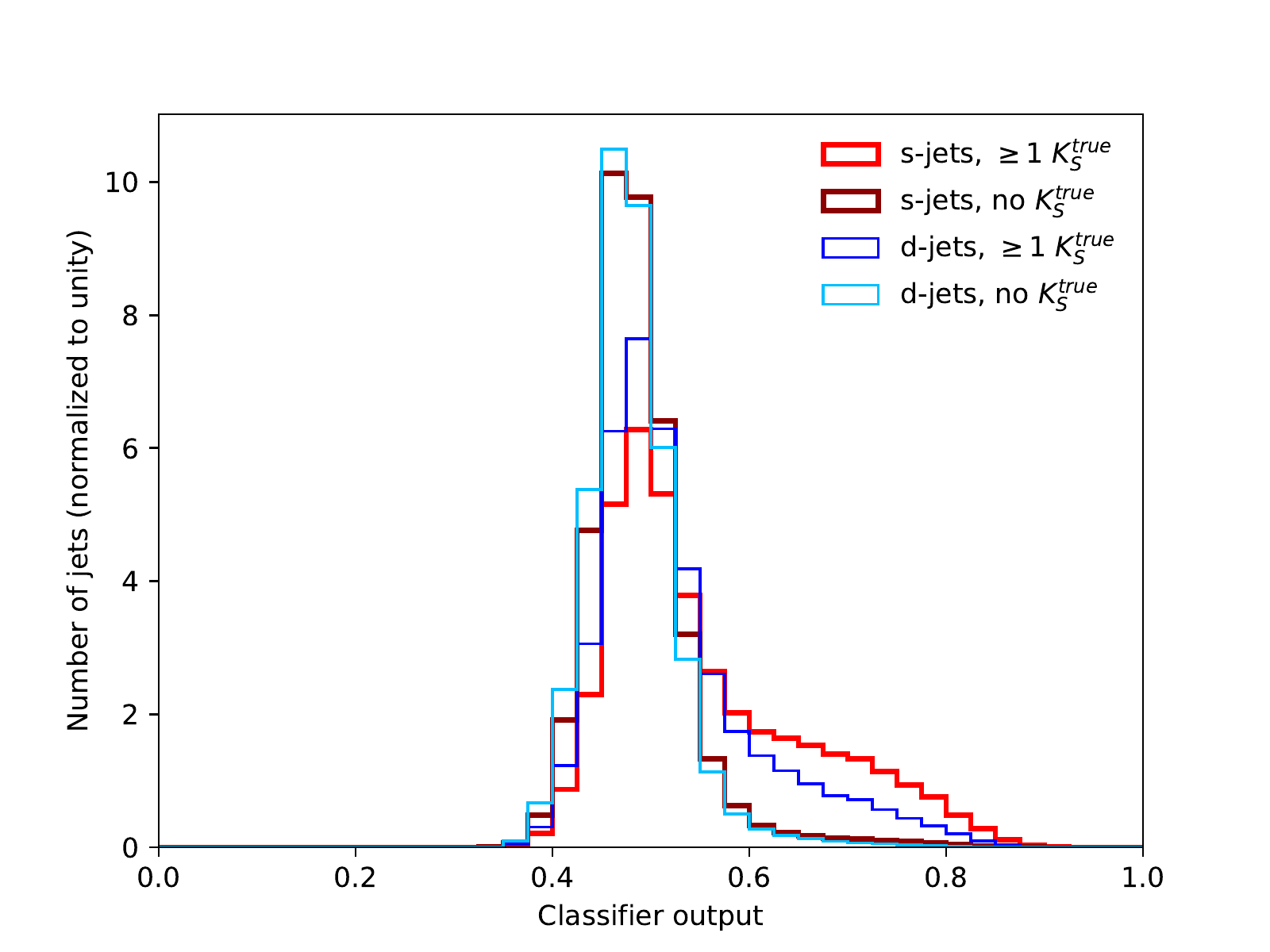}}\\
  \subfloat[]{\includegraphics[width=0.49\textwidth]{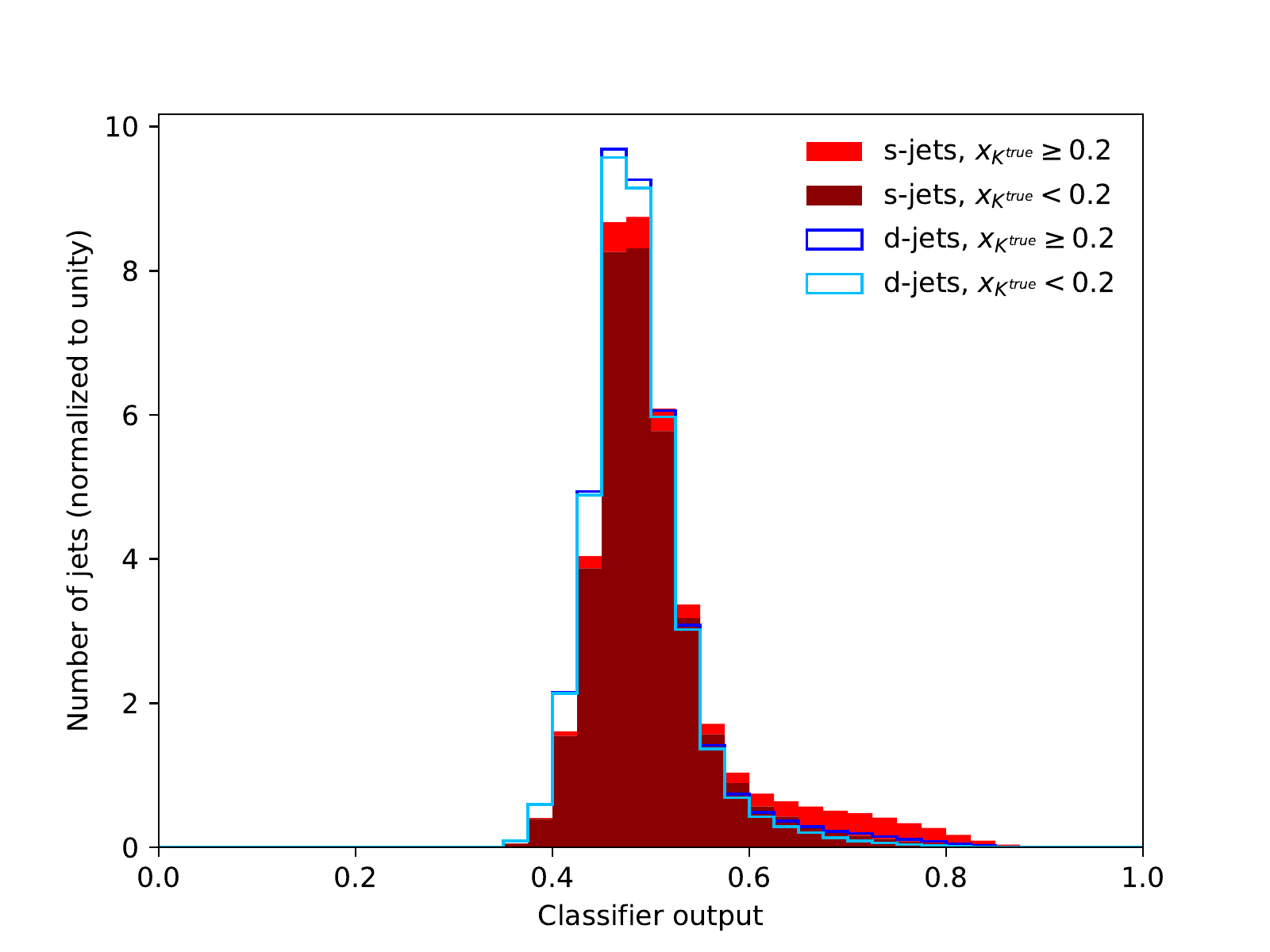}}
  \subfloat[]{\includegraphics[width=0.49\textwidth]{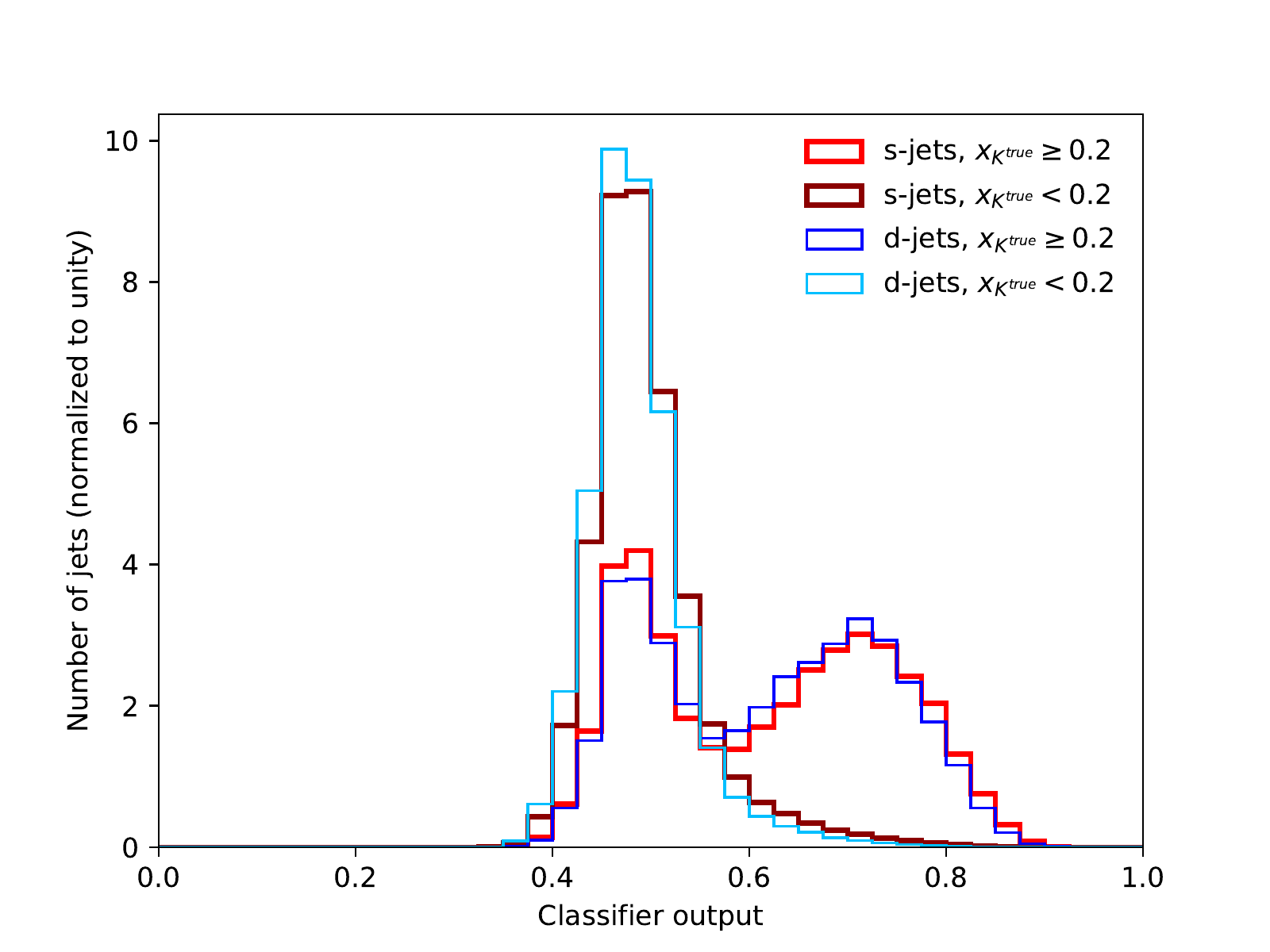}}\\
  \caption{
    Distribution of the classifier output evaluated on the test sample for $s$- (red) and $d$-jets (blue) for the network with 50 LSTM units and 3 fully-connected layers with 200, 100 and 50 nodes trained with all jets. The distributions are split (a,b) by requiring at least one true $K_S\rightarrow\pi^+\pi^-$ decay or no such decay or (c,d) by requiring $x_K$ of the true $K_S$ to be either larger or smaller than 0.2 (jets without $K_S\rightarrow\pi^+\pi^-$ decays are counted with an $x_K$ of 0). In figures (a) and (c), the split distributions are stacked on top of each other for $s$- and $d$-jets separately. In figures (b) and (d), the split distributions are shown separately and are each normalised to unity.
  }
  \label{fig:inv3}
\end{figure}

\begin{figure}
  \centering
  \subfloat[]{\includegraphics[width=0.49\textwidth]{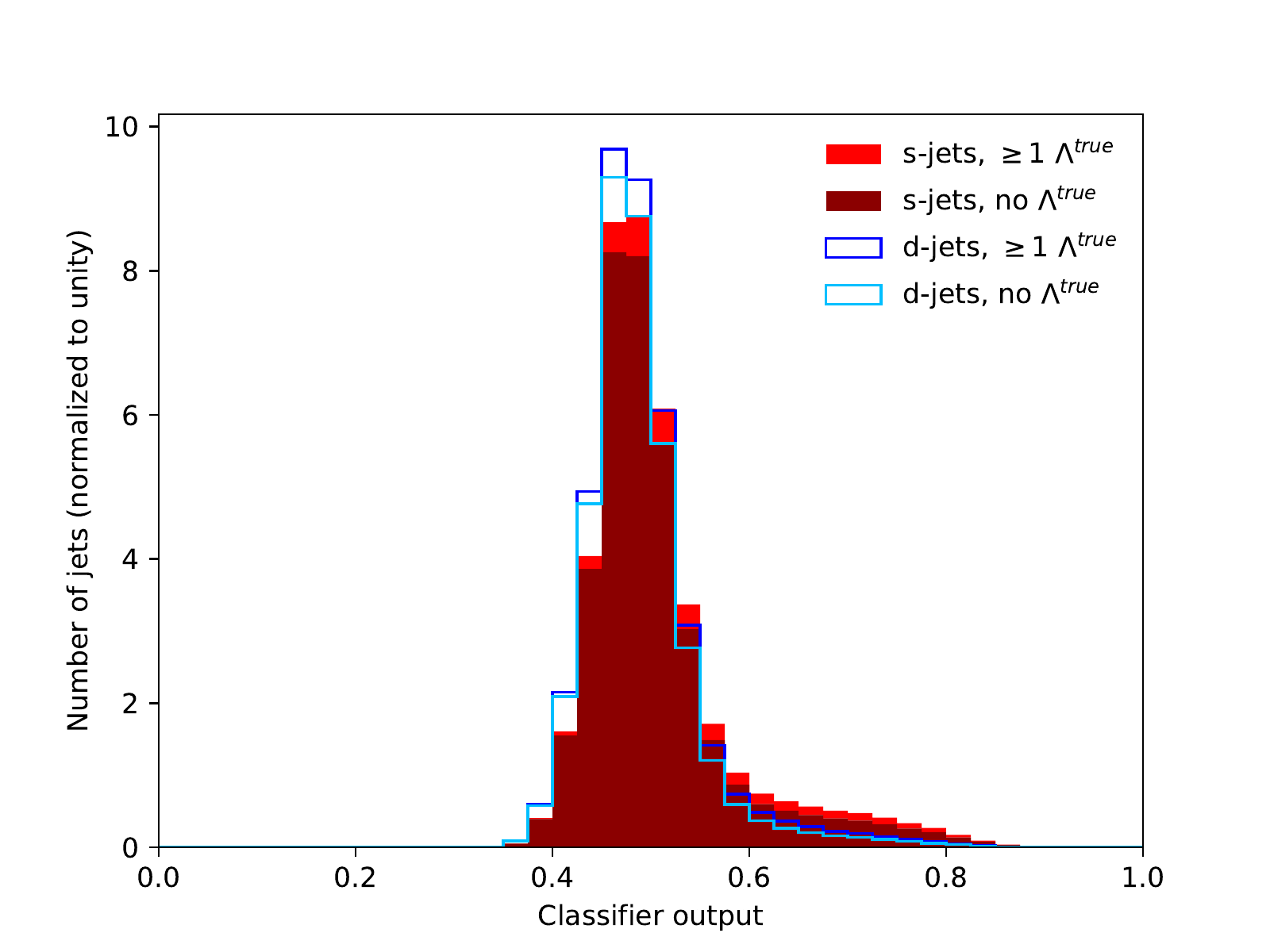}}
  \subfloat[]{\includegraphics[width=0.49\textwidth]{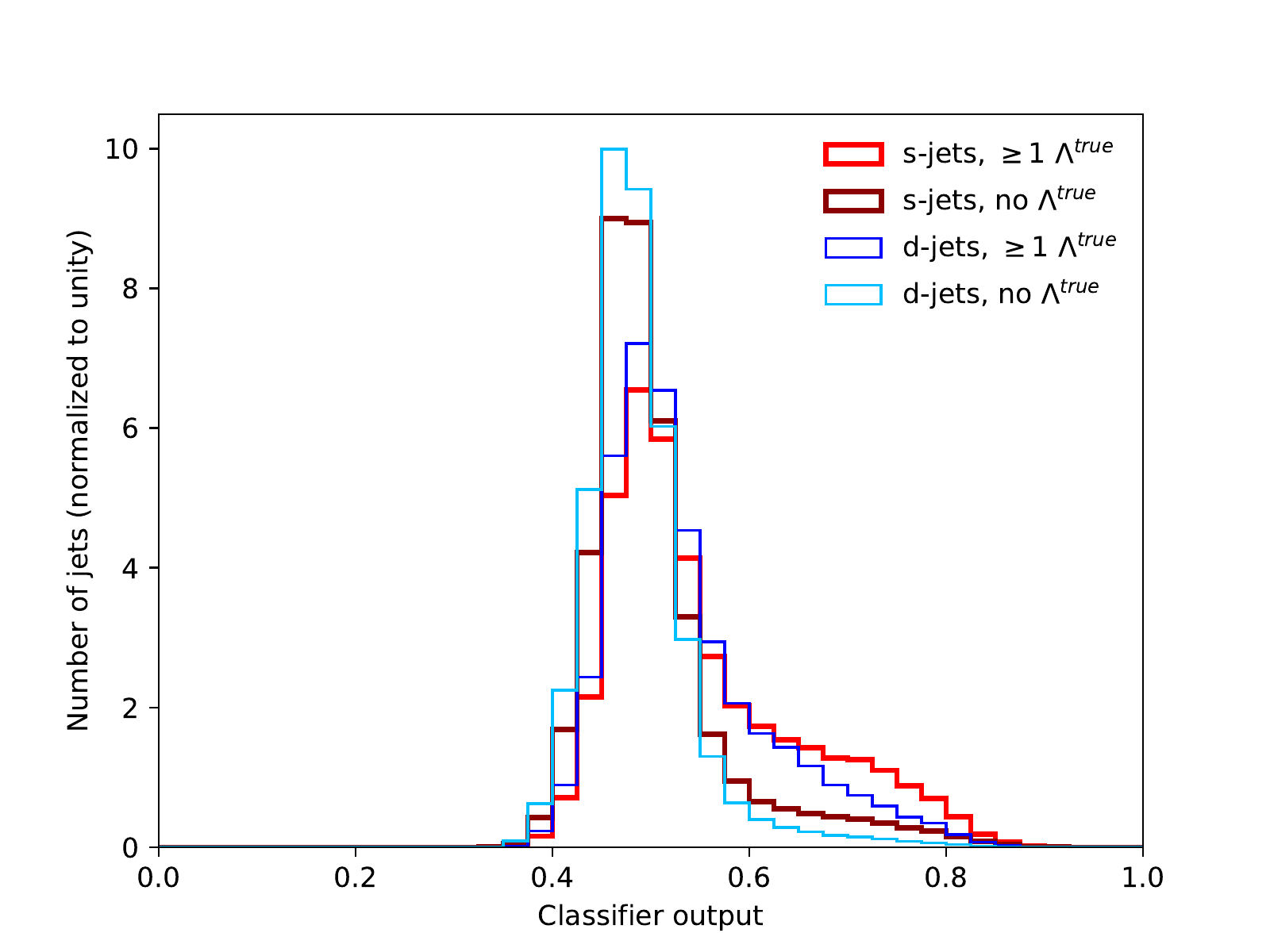}}\\
  \subfloat[]{\includegraphics[width=0.49\textwidth]{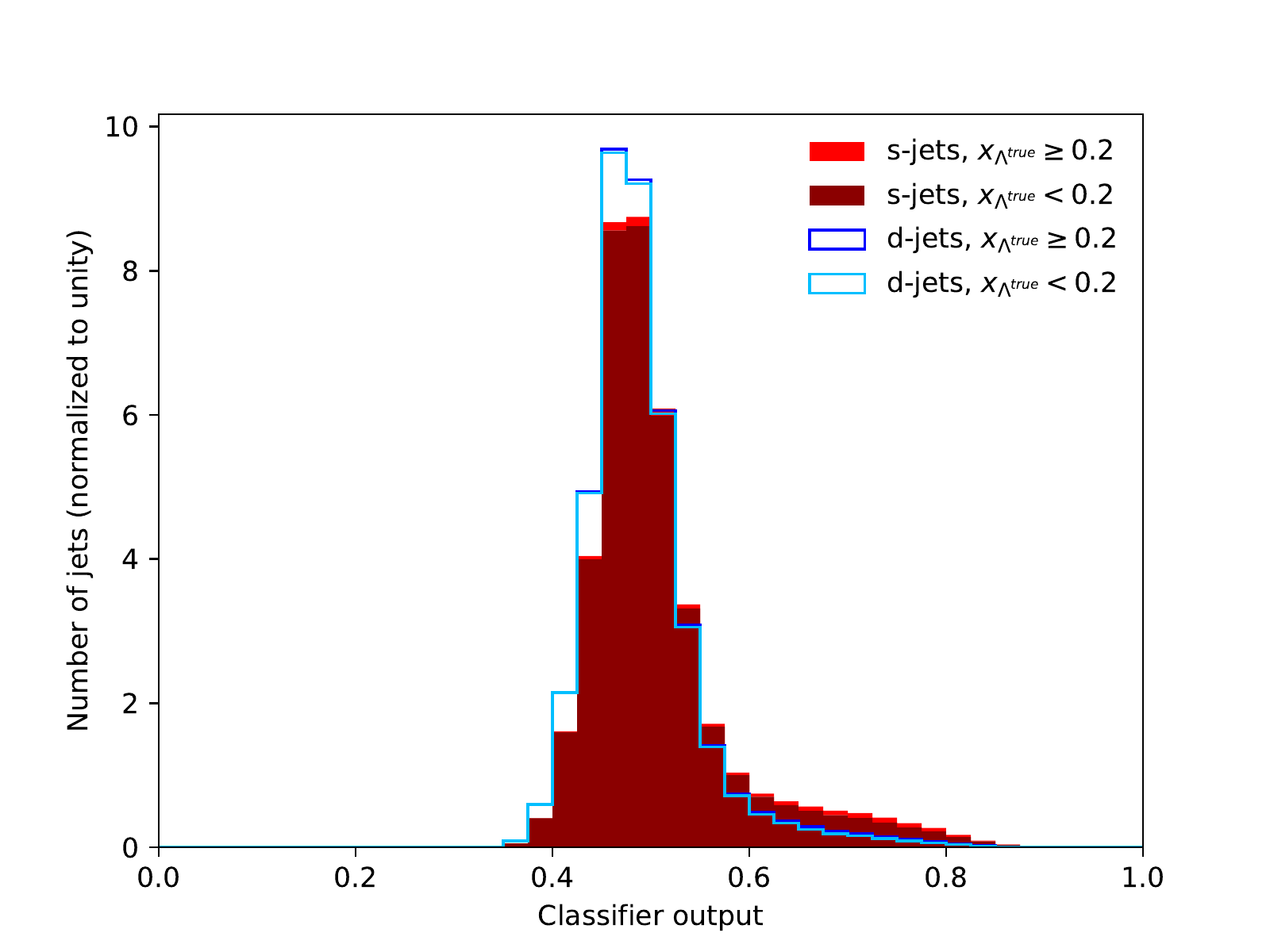}}
  \subfloat[]{\includegraphics[width=0.49\textwidth]{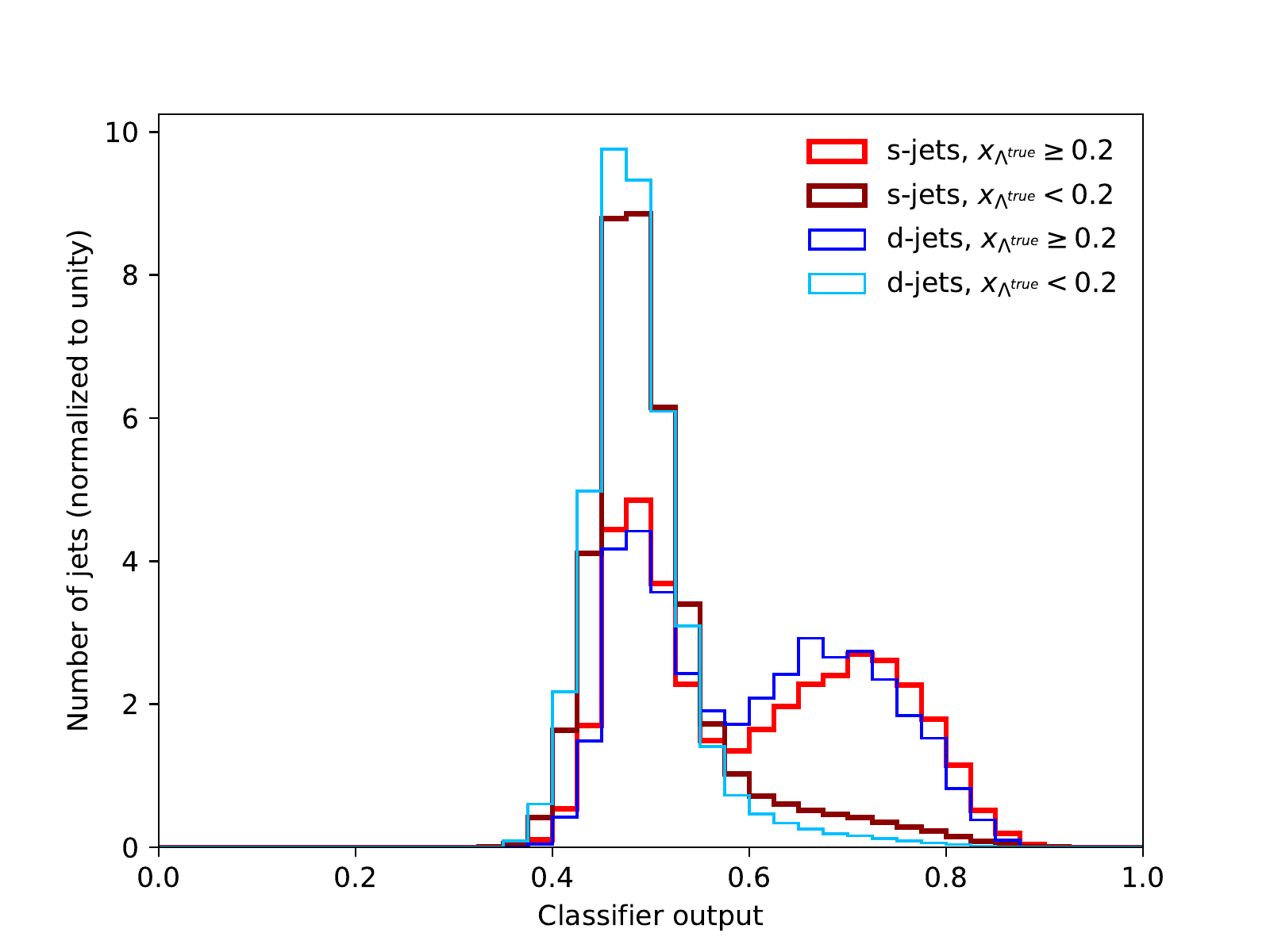}}\\
  \caption{
    Distribution of the classifier output evaluated on the test sample for $s$- (red) and $d$-jets (blue) for the network with 50 LSTM units and 3 fully-connected layers with 200, 100 and 50 nodes trained with all jets. The distributions are split (a,b) by requiring at least one true $\Lambda\rightarrow p\pi^-$ decay or no such decay or (c,d) by requiring $x_\Lambda$ of the true $\Lambda$ to be either larger or smaller than 0.2 (jets without $\Lambda\rightarrow p\pi^-$ decays are counted with an $x_\Lambda$ of 0). In figures (a) and (c), the split distributions are stacked on top of each other for $s$- and $d$-jets separately. In figures (b) and (d), the split distributions are shown separately and are each normalised to unity.
  }
  \label{fig:inv4}
\end{figure}

As the performance of the different methods for efficiencies smaller than 19\%, i.e. in the efficiency range of the $x_K$\&$x_{\Lambda}$ method, is similar, this suggests that the main discriminatory power between $s$-jets and $d$- and $u$-jets lies in the decays of neutral strange hadrons. Additional tracking information does not result in a better discrimination. However, the similarity of the performance also implies that the LSTM networks learn the major discriminatory features from these decays from the track series without that features of the reconstructed hadrons themselves would have been used in the training. It is hence interesting to investigate the correlations of the classifier output with variables that describe the presence of $K_S$ meson and $\Lambda$ baryon decays to either two charged pions or a proton and a charged pion. For this purpose, the classifier output distribution for the LSTM network from Section~\ref{sec:alltracks} (all jets) is split according to certain requirements in Figures~\ref{fig:inv1}--\ref{fig:inv4}. In each of these figures on the left side, the $s$- and $d$-jet distributions are decomposed according to the requirement. On the right side, the corresponding distributions are shown separately, each normalised to unity.

In Figure~\ref{fig:inv1}, the classifier output distribution is split by jets that contain at least one track that does not originate from the PV and jets with no such track. Jets with such a track are assigned higher classifier outputs than the other jets. While the classifier can discriminate between $s$-jets and $d$-jets that contain such a track, only small discrimination is achieved for the other jets. This indicates that the network indeed learns from the presence of tracks that do not originate from the PV, as expected from decays of strange hadrons.

In Figure~\ref{fig:inv2}~(a,b), the classifier output distribution is split by jets that contain at least one reconstructed $K_S$ candidate as described in Section~\ref{sec:Kshort}. Jets that contain a $K_S$ candidate are assigned higher classifier outputs than jets without a $K_S$ candidate. In Figure~\ref{fig:inv2}~(c,d), the classifier output is further split by requiring that the value of $x_K$ of the $K_S$ candidate must be larger or smaller than 0.2, where jets without a $K_S$ candidates are assigned an $x_K$ value of 0. Jets with a $K_S$ candidate with a large value of $x_K$ are assigned high values of the classifier output, and little discrimination remains between $s$- and $d$-jets. This indicates that the network learns from the presence of tracks that originate from the decay of $K_S$ mesons and it further learns that $K_S$ decays with a large value of $x_K$ are more likely to appear in $s$-jets than in $d$-jets.

Similar conclusions can be drawn from Figure~\ref{fig:inv3}, where the classifier output distribution is not split by the presence of reconstructed $K_S\rightarrow\pi^+\pi^-$ decays but by the presence of true $K_S\rightarrow\pi^+\pi^-$ decays obtained from the Monte Carlo truth record. The true $K_S$ mesons are geometrically matched to the reconstructed jet with the same $\Delta R$ criterion as used for the reconstructed $K_S$ meson decays. Analogously to Figure~\ref{fig:inv2}, the classifier output is split by the presence of a true $K_S$ meson decay~(a,b) and by the presence of such a decay with $x_K > 0.2$~(c,d). The same distributions are shown in Figure~\ref{fig:inv4} for true $\Lambda$ baryon decays to $p\pi^-$ and the same conclusions hold as for the $K_S$ mesons. When the classifier output is split by the presence of true strange hadrons (Figures~\ref{fig:inv3}~(a,b) and~\ref{fig:inv4}~(a,b)), lower values of the output are assigned to some jets that contain a true hadron, which is expected because one or both tracks of the hadron decay may not be reconstructed. Consequently, also Figures~\ref{fig:inv3}~(c,d) and~\ref{fig:inv4}~(c,d) show a two-peak structure for jets with a value of $x_K$ (or $x_\Lambda$) larger than 0.2. Overall, these studies indicate that the LSTM network learns from the presence of $K_S$ meson and $\Lambda$ baryon decays. As the maximal efficiency that the network trained with jets that have at least one track not from the PV is much larger than the maximal efficiency of the $x_K$\&$x_{\Lambda}$ method, this increase in efficiency reach cannot be only attributed to reconstructed $K_S$ and $\Lambda$ decays but must also be due to jets in which one of the tracks from the decay was lost. In contrast to the $x_K$\&$x_{\Lambda}$ method, the LSTM network is able to discriminate between $s$- and $d$- and $u$-jets also for jets that contain only partial decay information.

%% file: conclusions.tex
An algorithm for the identification of strange jets was presented that uses tracking information and is based on long short-term memory recurrent neural networks. The algorithm was compared to a simple benchmark algorithm, which uses the fraction of the jet transverse momentum that is carried by reconstructed $K_S\rightarrow\pi^+\pi^-$ and $\Lambda\rightarrow p\pi^-$ decays. The signal efficiency of the benchmark algorithm is limited to values smaller than 19\%. In this range of efficiencies, the neural-network-based algorithm achieves a similar performance as the benchmark algorithm. However, it allows for the identification of strange jets also at higher efficiencies. For signal efficiencies of 30\% and 70\%, background efficiencies of 22\% and 64\% are achieved, respectively. An average pile-up of 35 interactions per bunch-crossing was used in the training. Comparing to a training without pile-up interactions, it was shown that the algorithm's performance is not drastically deteriorated by the presence of pile-up tracks. Although trained on jets that originate from the hadronisation of down quarks as background, the performance is very similar for jets that originate from the hadronisation of up quarks. Investigations of the network output indicate that the neural network learns a large part of its discrimination power from the presence of strange-hadron decays from the track pattern without that features of such reconstructed strange-hadron decays are used as input to the network.

Strange-tagging algorithms complement the existing algorithms for flavour tagging. Although the discrimination of $s$-jets and $d$- and $u$-jets is not very strong, strange tagging may open new research possibilities at colliders, notably at the LHC. Possible applications of strange tagging are the search for the decay of the Higgs boson to strange quarks, the measurement of the decay $t\rightarrow W^+ s$ and the search for new particles with a significant branching ratio to final states with strange quarks.

While in this study a simplified detector simulation was used, more refined simulations would be used for the training at an experiment, which would for example account for the presence of fake tracks, which were not considered in this study. The simulated efficiencies of the strange-tagging algorithm could be calibrated using data from $W^+\rightarrow c\bar{s}$ and $W^-\rightarrow \bar{c}s$ decays in top-antitop-quark production. While the simple algorithm based on $K_S$ and $\Lambda$ reconstruction---despite its low efficiency---may readily be used at an experiment, already the presence of tracks that do not originate from the primary vertex provides a distinct signature and the prospects for the use of long short-term memory may be promising. The long short-term memory network that uses all jets as input, however, may need to be studied with respect to the accuracy of the modelling in Monte Carlo simulations and the associated systematic uncertainties. The strategy proposed in this work, which is based only on tracking information, may be further improved by including information from the calorimeter and muon systems in order to explore features in the energy-distribution pattern and the presence of in-flight decays of $K^\pm$ mesons.